\documentclass[journal=inoraj,manuscript=article]{achemso}
\usepackage[version=3]{mhchem} % Formula subscripts using \ce{}
\usepackage[T1]{fontenc} % Use modern font encodings
\author{Fabiana Machado Ferreira De Araujo}
\affiliation{Institute of Physics, Carl-von-Ossietzky Universit{\"a}t Oldenburg, 26129 Oldenburg, Germany}
\altaffiliation{Contributed equally to this work}
\author{Daniel Duarte-Ruiz}
\affiliation{Institute of Physics, Carl-von-Ossietzky Universit{\"a}t Oldenburg, 26129 Oldenburg, Germany}
\altaffiliation{Contributed equally to this work}
\author{Holger-Dietrich Sa{\ss}nick}
\affiliation{Institute of Physics, Carl-von-Ossietzky Universit{\"a}t Oldenburg, 26129 Oldenburg, Germany}
\author{Marie C. Gentzmann}
\altaffiliation{Present address: Bundesanstalt für Geowissenschaften und Rohstoffe (BGR), Stilleweg 2, 30655, Hannover, Germany}
\affiliation{Bundesanstalt für Materialforschung und –prüfung (BAM), Unter den Eichen 87, 12205, Berlin, Germany}
\author{Thomas Huthwelker}
\affiliation{Paul Scherrer Institut, Swiss Light Source (SLS), 5232, Villigen, Switzerland}
\author{Caterina Cocchi}
\affiliation{Institute of Physics, Carl-von-Ossietzky Universit{\"a}t Oldenburg, 26129 Oldenburg, Germany}
\alsoaffiliation{Physics Department and IRIS Adlershof, Humboldt-Universit\"at zu Berlin, 12489 Berlin, Germany}
\email{caterina.cocchi@uni-oldenburg.de}

\title  {Electronic Structure and Core Spectroscopy of Scandium Fluoride Polymorphs}

\begin{document}

%\begin{tocentry}
%
%\includegraphics[height=4.45 cm]{toc.png}
%
%\end{tocentry}


\newpage

\begin{abstract}
The microscopic knowledge of the structural, energetic, and electronic properties of scandium fluoride is still incomplete, despite the relevance of this material as an intermediate for the manufacturing of Al-Sc alloys.
In a work based on first-principles calculations and x-ray spectroscopy, we assess the stability and the electronic structure of six computationally predicted \ce{ScF3} polymorphs, two of which correspond to experimentally resolved single-crystal phases.
In the theoretical analysis based on density-functional theory (DFT), we identify similarities among the polymorphs based on their formation energies, charge-density distribution, and electronic properties (band gaps and density of states).  
We find striking analogies between the results obtained for the low- and high-temperature phases of the material, indirectly confirming that the transition occurring between them mainly consists of a rigid rotation of the lattice.
With this knowledge, we examine the x-ray absorption spectra from the Sc and F K-edge contrasting first-principles results obtained from the solution of the Bethe-Salpeter equation on top of all-electron DFT with high-energy-resolution fluorescence detection measurements.
The analysis of the computational results sheds light on the electronic origin of the absorption maxima and provides information on the prominent excitonic effects that characterize all spectra.
Comparison with measurements confirms that the sample is mainly composed of the high- and low-temperature polymorphs of \ce{ScF3}.
However, some fine details in the experimental results suggest that the probed powder sample may contain defects and/or residual traces of metastable polymorphs.

\end{abstract}

\newpage
\section{Introduction}
The highly valuable transition metal Scandium (Sc) is suitable for various high-tech applications in solid oxide fuel cells, Al-Sc alloys, and laser industry~\cite{iver+01jecs,toro+17book,dori+18book}.  Due to its dispersive nature, it is currently recovered as a by-product of titanium dioxide, zirconium dioxide, uranium, and nickel production~\cite{grad21}. In addition, Sc recovery from bauxite residues is under investigation or implemented on a pilot plant scale in several locations~\cite{petr+16book,balo+21book,gent+22jge}.
The ionic crystal scandium fluoride (\ce{ScF3}) is an important intermediate compound for the manufacturing of Al-Sc alloys where it is preferred over scandium oxide (\ce{Sc2O3})~\cite{pete+19jsm}. The use of Sc in Al-alloys strongly enhances the properties of the material since it positively affects grain refinement, precipitation hardening, and superplasticity, and increases recrystallization and corrosion resistance. \ce{ScF3} can be produced by solvent extraction from an Sc-containing solution followed by stripping and precipitation in form of a hydroxide or oxalate salt. 
The precipitate is first calcined to obtain \ce{Sc2O3} and subsequently fluorinated with hydrofluoric acid (HF) to finally obtain \ce{ScF3}.
Alternative methods to directly obtain \ce{ScF3} without using HF have recently been developed~\cite{yagm+21mp}. 

From a fundamental viewpoint, \ce{ScF3} has received considerable interest in the last few years due to its negative thermal expansion, namely its remarkable ability to shrink when heated.\cite{grev+10jacs,li+11prl,laza+15prb,pisk+16prb,attf18fc,oba+19prm,dove+20prb,wei+20prl}
This property is closely related to the electronic structure of this material and the crystallographic arrangement of the two atomic species therein: Upon heating, F atoms oscillate around their bonds with Sc leading to an overall contraction of the crystalline volume.\cite{li+11prl}
The structural flexibility of this material is prone to polymorphism.
At low-temperature and low-pressure conditions, \ce{ScF3} crystallizes in a cubic lattice of space group $\text{Pm}\bar{\text{3}}\text{m}$ with hexacoordinated Sc atoms forming corner-sharing octahedra~\cite{grev+10jacs}.
At increasing temperature and/or pressure, the material undergoes a phase transition and assumes a trigonal configuration (space group $\text{R}\bar{\text{3}}\text{c}$) in which the octahedra are rotated around their axes.~\cite{alek+02jetp,grev+10jacs}

Beyond the above-cited studies on the peculiar structural and thermal properties of \ce{ScF3}, the knowledge of the fundamental properties of this material is still incomplete.
Questions regarding the stability of the crystal and its polymorphism, as well as the charge distribution within the lattice demand answers for a deeper understanding of the microscopic characteristics of \ce{ScF3}.
This information will provide not only further insight into the intrinsic properties of this material but, more generally, on the electronic structure of Sc-based compounds.
This fundamental knowledge is furthermore crucial for the development of applications that take advantage of the special properties of \ce{ScF3} such as its negative thermal expansion.

In this paper, we present a joint theoretical and experimental study to get insight into the structure-property relationships in scandium fluoride.
In the framework of density-functional theory (DFT), we investigate six \ce{ScF3} polymorphs, including the experimentally resolved low- and high-temperature phases~\cite{alek+02jetp,grev+10jacs} as well as four computationally predicted structures~\cite{jain+13aplm}, and perform a systematic analysis evaluating bond lengths, formation energies, partial charges, and electronic properties.
These results are the baseline to interpret x-ray absorption spectra measured from the Sc and F K-edge on a single powder sample of \ce{ScF3}.
The analysis of the spectra computed from first principles by solving the Bethe-Salpeter equation (BSE) enables relating spectral fingerprints with the electronic structure of the material polymorphs and it provides valuable information regarding the excitonic effects.
A systematic comparison between the computed spectra and the measurements suggests that the powder sample probed in the experiments is predominantly made of the known, stable phases of the material.
The spectroscopic analysis of the polymorphs that are to date only computationally predicted enriches the established knowledge on \ce{ScF3} and provides the community with additional insight into the electronic structure of this compound in relation to the stability, the charge distribution, and the electronic properties of its different crystal structures.

%%%%%%%%%%%%%%%%%%%%%%%%%%%%%%%%%%%%%
\section{Methodology}
\subsection{Theoretical background and computational details}

The six structures considered in the computational analysis are taken from Materials Project~\cite{jain+13aplm} (see Supporting Information, Table~S1) and they are not further relaxed.
The ground-state properties of the considered \ce{ScF3} polymorphs are calculated from DFT using the plane-wave, pseudopotential code Quantum ESPRESSO.\cite{gian+17jpcm}
In all cases, the Brillouin zone is sampled by a homogeneous 8$\times$8$\times$8 \textbf{k}-grid.
To ensure converged results, a cut-off value of 150~Ry and 1200~Ry for the plane waves and the density are chosen, respectively.
The projector augmented-wave method is applied to account for the core electrons using pseudopotentials from the \texttt{pslibrary}~\cite{dalc14cms}.
The exchange-correlation potential is expressed in the generalized-gradient approximation using the Perdew-Burke-Ernzerhof (PBE) parametrization~\cite{perd+96prl}.
Bader charge analysis is performed to evaluate the electron-density distribution and hence to harvest information about the character of the chemical bond. For this task, the code developed by Henkelmann \textit{et al.}~\cite{henk+06cms,sanv+07jcc,tang+09jpcm} is used.

X-ray absorption spectra are calculated from first principles through the solution of the BSE within the all-electron and full-potential framework implemented in the code \texttt{exciting}~\cite{gula+14jpcm}, which grants direct access to core electrons as the initial transition states~\cite{vorw+17prb,vorw+19es}.
In practice, the BSE~\cite{salp-beth51pr}, which is the equation of motion for the electron-hole correlation function, is solved as an effective, time-independent two-particle Schr\"odinger equation~\cite{vorw+17prb},
\begin{equation}\label{eq:bse}
    \sum_{c'u'\textbf{k'}}\hat{H}^{BSE}_{cu\textbf{k},c'u'\textbf{k'}}A^{\lambda}_{c'u'\textbf{k'}}=E^{\lambda}A^{\lambda}_{cu\textbf{k}},
\end{equation}
where $c$ and $u$ stand for core and unoccupied states, respectively.
The Hamiltonian in Eq.~\eqref{eq:bse} is composed of three terms: $\hat{H}^{BSE} = \hat{H}^{diag} + \hat{H}^{dir} + \hat{H}^x$.
The first one, $\hat{H}^{diag}$, represents vertical transitions from core to unoccupied levels; the second one, $\hat{H}^{dir}$, accounts for the electron-hole Coulomb attraction and includes the statically screened Coulomb potential; the third one, $\hat{H}^x$, is the exchange interaction: this term is repulsive due to the opposite charge of electron and hole. For further details on the BSE Hamiltonian we refer the readers to the specialized literature~\cite{rohl-loui00prb,pusc-ambr02prb,onid+02rmp}.
Neglecting the last two terms, namely solving Eq.~\eqref{eq:bse} for $\hat{H}^{BSE} = \hat{H}^{diag}$, corresponds to the so-called independent-particle approximation (IPA).
Eigenvalues and eigenvectors of Eq.~\eqref{eq:bse} represent excitation energies and excited states, respectively, and both enter the expression of the imaginary part of the macroscopic dielectric function which is commonly adopted to represent absorption spectra:
\begin{equation}\label{eq:epsilon}
    \Im{\varepsilon_{M}}=\frac{8\pi^{2}}{\Omega}\sum_{\lambda}\left|\sum_{cu\textbf{k}}A^{\lambda}_{cu\textbf{k}}\dfrac{\langle c|\hat{\textbf{p}}|u\textbf{\textbf{k}}\rangle}{\epsilon_{u\textbf{k}} - \epsilon_{c}+\Delta}\right|^{2}\delta(\omega-E_{\lambda}).
\end{equation}
In the square modulus of Eq.~\eqref{eq:epsilon}, we recognize the transition matrix elements for the momentum operator $\hat{\textbf{p}}$, where in the denominator, a scissors operator $\Delta$ is added to mimic the quasi-particle correction to the core and the conduction levels. For a one-to-one comparison of each computed spectrum with the experimental reference, we choose different values of $\Delta$ taken with respect to the most intense resonance in the measured energy window. 
To this end, $\Delta$=105.8~eV for polymorph (1), $\Delta$=107.6~eV for (2), $\Delta$=106.4~eV for (3), $\Delta$=106.1~eV for (4), $\Delta$=105.9~eV for (5), and $\Delta$=102.1~eV for (6) are taken for the Sc K-edge spectra. 
For the F K-edge spectra, chosen values are $\Delta$=33.0~eV (1), $\Delta$=33.8~eV (2), $\Delta$=33.3~eV (3), $\Delta$=33.9~eV (4), $\Delta$=33.1~eV (5) and $\Delta$=32.5~eV (6). The unit cells of the six considered polymorphs are sketched in Figure~\ref{fgr:Dis} and corresponding crystallographic information are reported in Table~\ref{tbl:formE}.

In the underlying DFT calculations, performed with \texttt{exciting} using the PBE functional, a 4$\times$4$\times$4 \textbf{k}-mesh is used to sample the Brillouin zone of all structures except for polymorph (6), for which a 6$\times$6$\times$6 \textbf{k}-grid is employed. 
Muffin tin radii for Sc atoms are chosen as large as 1.90~bohr (polymorphs 1 and 2), 1.97~bohr (3), 1.88~bohr (4), 1.92~bohr (5), 1.89~bohr (6) and as large as 1.71~bohr (1 and 2), 1.77~bohr (3), 1.69~bohr (4), 1.73~bohr (5), 1.69~bohr (6) for F atoms.
The cutoff value R$_{MT}$G$_{MAX}=8$ is adopted for all systems.
In the BSE calculations, a $\Gamma$-shifted \textbf{k}-mesh with 4$\times$4$\times$4 points is employed for all polymorphs except for (3), for which a 6$\times$6$\times$6 \textbf{k}grid is taken.
The screened Coulomb potential is computed from the random-phase approximation including 40 (polymorph 1), 30 (polymorph 2), 15 (polymorph 3), 20 (polymorph 4), 60 (polymorph 5), and 90 (polymorph 6) empty bands.
Local-field effects are accounted for by choosing energy cutoffs of 108.85~eV (polymorph 1), 81.63~eV (polymorph 2), 163.27~eV (polymorph 3), 108.85~eV (polymorph 4), 54.42~eV (polymorphs 5 and 6).
We include transitions from core levels to conduction bands within an energy range of 37~eV above the conduction band minimum. 

%%%%%%%%%%%%%%%%%%%%%%
\subsection{Experimental Methods}
\ce{ScF3} was provided by the company KBM in the framework of the Horizon 2020 project SCALE. It originates from a Chinese manufacturer and was tested for purity with inductively coupled plasma optical emission spectroscopy and microscopy~\cite{gent+21ag}.  
X-ray absorption near edge structure (XANES) spectra were taken at the PHotons for the Exploration of Nature by Imaging and XAFS (PHOENIX I) undulator beamline of the Swiss Light Source, Paul-Scherrer-Institute (PSI), Villigen, Switzerland.

The Sc K-edge spectra were measured at the PHOENIX I branch line, which covers an energy range from 0.8 to 8~keV, using a double crystal monochromator. 
To generate monochromatic light at the Sc K edge, a Si(111) crystal was employed, providing an energy bandwidth for the incoming photons of 0.4-0.5~eV.
The powder sample of \ce{ScF3} was pressed into a pellet and XANES spectra were taken under vacuum (ca. 10$^{-5}$-10$^{-6}$~mbar), by scanning the energy of the incoming photons over a range from 4400 to 4600~eV and recording the fluorescent light using an energy dispersive silicon drift detector (4 element Vortex detector, manufacturer Hitachi). The beamsize for the powder sample was 1.5$\times$1.5~mm. The absolute flux impinging the sample is in the order of 10$^{10}$-10$^{11}$ photon/s although not quantitatively measured for each measurement. For the normalization of the XANES spectra, 
the incoming flux, $I_0$, was taken from a total electron yield (TEY) signal measured on a Ni-coated polyethylene terephthalate foil, located about 1~m upstream of the sample in a vacuum chamber, which is held at $\sim$10$^{-7}$~mbar. 
Additionally, High Energy Resolution Fluorescence Detected (HERFD) spectra were taken using a new compact von Hamos spectrometer implemented in the end station. 
Briefly, the spectrometer uses a segmented Si(111) crystal (radius 7~cm) which is mounted in backscattering geometry. The main axis of the spectrometer is vertically under 90$^{\circ}$ relative to the incoming beam. The fluorescent photons are collected on a novel 2D Moench detector~\cite{rami+17jinst}, which is mounted on a 2D manipulator to align the in-vacuum spectrometer.  The spectrometer is operated with a microfocussed beam of 10$\times$50~$\mu$m. Therefore, despite its small crystal radius, it provides an energy resolution of about 0.5~eV, which is sufficient for emission spectroscopy at tender x-rays. 
The energy resolution of the spectrometer was derived from the experimentally determined bandwidth of $\sigma \approx 0.5$~eV of elastically scattered photons. This number is a convolution of the true photon bandwidth of 0.3-0.4~eV and a similar spectrometer resolution, both given by the rocking curve of the Si(111) crystal. 

For each excitation energy, the Sc $K_{\alpha}$ fluorescence line is derived from the 2D image taken with the Moench detector. To derive the HERFD spectra, only the central part of the $K_{\alpha 1}$ emission lines is integrated for each excitation energy (see Supporting Information for further details).
The integrated width of about 0.9~eV, which is approximately three times the von Hamos spectrometer energy resolution. To obtain the spectra, the excitation energy was chosen with a step of 0.3~eV around the edge, and 0.5 to 3~eV in the pre- and post-edge regions was chosen. 

To measure x-ray absorption spectra at the F K-edge, the PHOENIX II endstation was used. This endstation is attached to the X-Treme beamline~\cite{piam+12jsr} which shares the undulator with PHOENIX. Here, monochromatic light is generated using a planar grating monochromator (energy resolution < 0.2 eV).  
Data were taken in both TEY mode and in fluorescence mode using a 1-element energy dispersive silicon drift diode (manufacturer Ketek). 
The beam was shaped as a round spot of 2~mm in diameter. The photon flux on the sample is on the order of 10$^{10}$-10$^{11}$ photon/s but it was not measured prior to this experiment.
All samples were measured in a vacuum chamber kept at about 10$^{-6}$~mbar. It is separated from the beamline vacuum by a 0.5~$\mu$m thin silicon nitride window. The experiments were taken in both fluorescence and TEY mode.
The photon flux, $I_0$, hitting the sample was measured separately as a TEY signal from a sample-free part of the copper sample holder. This signal was then used to normalize the data.
Fundamentally, $I_0$ could be measured simultaneously by a partially transparent device, such as a gold mesh or a Ni-coated foil, upstream of the sample. At soft energies, such a device would absorb a significant fraction of the photons. This will hamper the $I_0$ measurement, as the photon flux on the $I_0$ device will differ from the one on the sample, and as the elemental composition of the device will introduce an unwanted signature to the $I_0$ measurement. Such pitfalls can be excluded by taking $I_0$ from a Cu plate inserted into the sample holder. The approach of subsequent $I_0$ measurement is only possible, because the photon source, a third-generation synchrotron, operates in top-up mode and has sufficient flux stability of less than 1\%, as needed for proper normalization recorded signal.

%%%%%%%%%%%%%%%%%%%%%%%%%%%%%%%%%%%%%

\section{Results and Discussion}
\subsection{Systems and Structural Properties}

\begin{figure}
 \centering
 \includegraphics[width=0.5\textwidth]{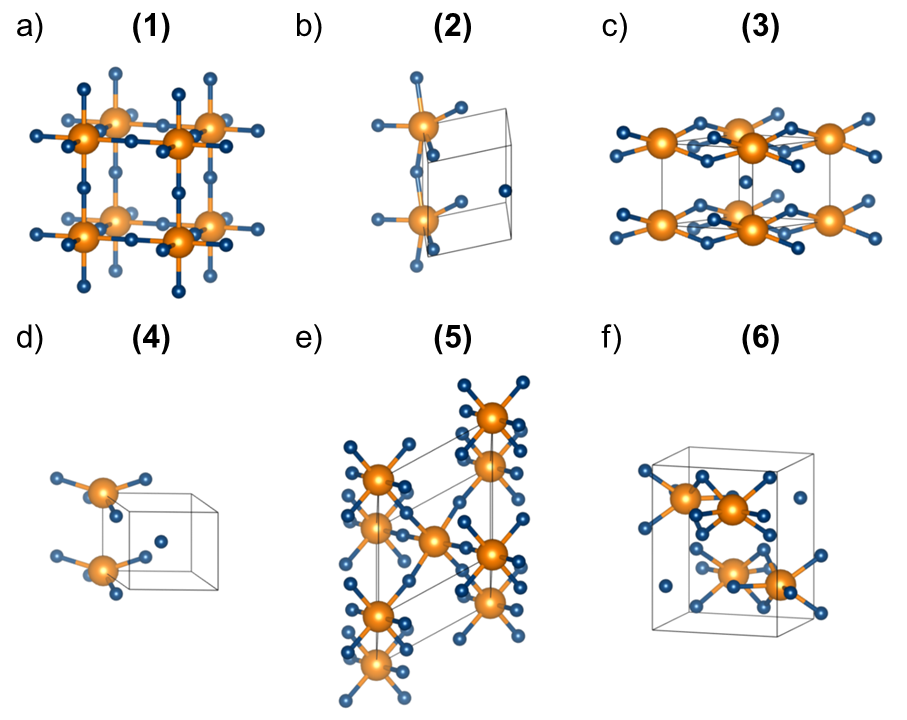}
 \caption{Ball-and-stick representation of the $\text{ScF}_\text{3}$ polymorphs considered in this work. a) Polymorph (1) with cubic lattice; b) polymorph (2) with monoclinic lattice; c) polymorph (3) with orthorhombic lattice; d) polymorph (4) with monoclinic lattice; e) polymorph (5) with trigonal lattice; f) polymorph (6) with orthorhombic lattice. Sc atoms are depicted in orange and F atoms in blue. Unit cells are marked by thin black lines; in panel a), the sides overlap with the Sc-F bonds. Plots produced with the visualization software VESTA~\cite{momm-izum11jacr}.}
 \label{fgr:Dis}
\end{figure}

To model \ce{ScF3}, we consider the six polymorphs visualized in Figure~\ref{fgr:Dis}; their structural and electronic characteristics are summarized in Table~\ref{tbl:formE}.
In addition to the low-temperature, low-pressure cubic crystal (1) and the high-pressure trigonal polymorph (5), two monoclinic phases (2 and 4) and two orthorhombic ones (3 and 6) are considered.
To the best of our knowledge, there is no experimental evidence for those structures (2, 3, 4, 6).
Yet, it is meaningful to include them in this study since the probed \ce{ScF3} powder is not assumed to contain monocrystalline particles. In fact, residuals of other polymorphs, although not common, cannot be excluded. 

\begin{table}[h!] 
\caption{Space group, lattice name, formation energy per atom ($E_{form}$ in eV/atom), as well as electronic ($E_{\text{gap}}$) and optical ($E_{\text{gap}}^{\text{opt}}$) gaps (both in eV) computed for the considered \ce{ScF3} polymorphs.}
  \label{tbl:formE}
    \centering
    \begin{tabular}{llllll}
    \hline
    Nr. & Space group & Lattice & $E_{form}$ & $E_{\text{gap}}$ & $E_{\text{gap}}^{\text{opt}}$ \\
    \hline
    (1) & $\text{Pm}\bar{\text{3}}\text{m}$ [221] & cubic &  -4.43 &  6.12 &  6.64 \\
    (2) & C2 [5] & monoclinic &  -4.19 &  5.28 &  5.52 \\
    (3) & C222 [21] & orthorhombic &  -3.56 &  1.84 &  2.31 \\
    (4) & C2 [5] & monoclinic &  -3.52 &  1.80 &  2.09 \\
    (5) & $\text{R}\bar{\text{3}}\text{c}$ [167] & trigonal &  -4.43 &  6.14 &  6.14 \\
    (6) & Pnma [62] & orthorhombic &  -4.32 &  5.77 &  5.81\\
    \hline
    \end{tabular}
    \label{tab:my_label}
\end{table}

\begin{figure}[h!]
 \centering
 \includegraphics[width=0.5\textwidth]{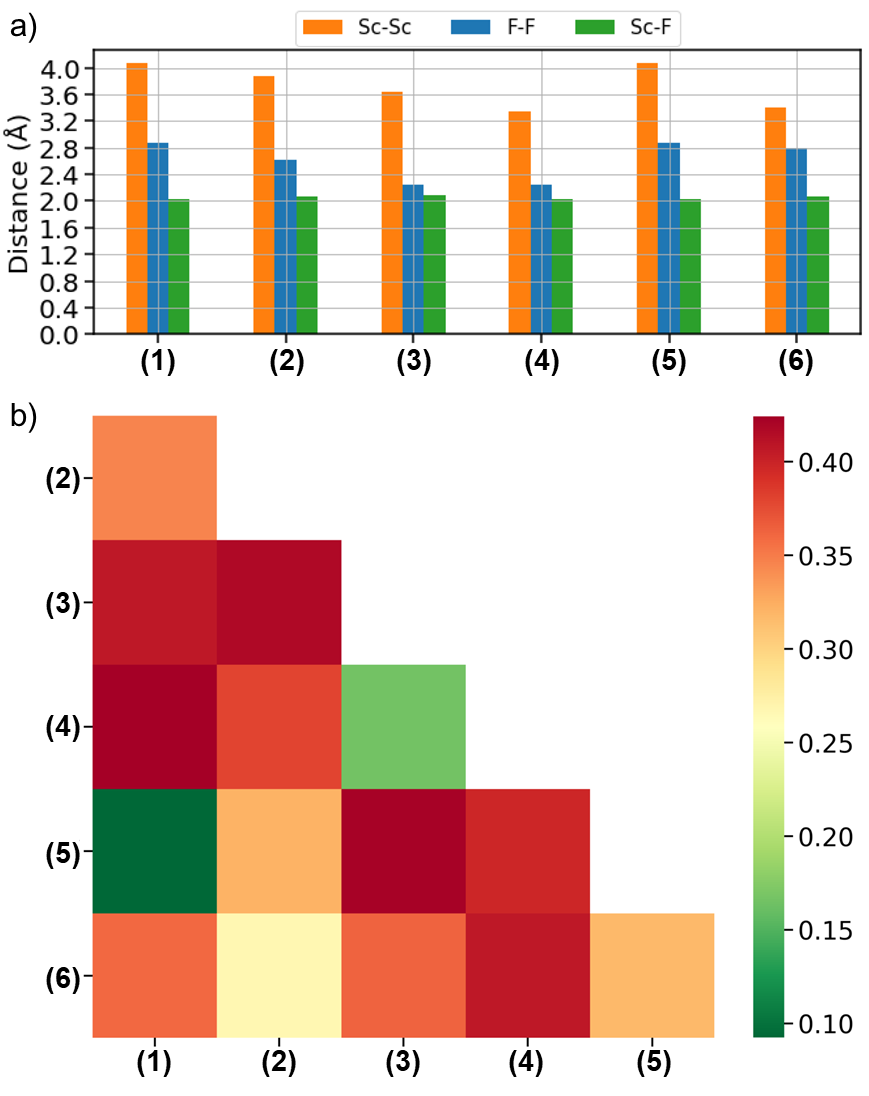}
 \caption{
 a) Absolute values of the averaged interatomic distances for the considered \ce{ScF3} polymorphs and b) similarity matrix of the averaged interatomic distances based on the pair-wise cosine-distances evaluated from the F-fingerprint metric: larger similarities correspond to low values (green) while smaller similarities to high values (red). 
 }
 \label{fgr:fingerprint}
\end{figure}

A quantitative analysis of the structural characteristics of the six considered \ce{ScF3} polymorphs displayed in Figure~\ref{fgr:Dis} reveals significant similarities among them.
First, we examine the mutual distances between Sc and F that are evaluated as the positions of the first peak in the F-fingerprint function~\cite{ogan-vall09jcp} of the element pair (see Figure~\ref{fgr:fingerprint}a).
Interestingly, we find that structures (1) and (5), namely the experimentally resolved low- and high-temperature phases of \ce{ScF3}, respectively, are characterized by almost identical interatomic distances, in line with the understanding that the pressure-induced phase transition corresponds to a rigid rotation of the octahedra in the lattice~\cite{grev+10jacs}.
The separation between pairs of Sc (F) atoms amounts to 4.1~\AA{} (2.9~\AA{}) while the Sc-F bond length is equal to 2.0~\AA{}.
The structural similarity between these two phases is also in line with the almost identical x-ray diffraction spectra of the two polymorphs (see Figure~S4 in the Supporting Information).
In the other polymorphs, this distance does not change significantly: this behavior can be understood considering the Sc-F bond as an intrinsic property of the compound, regardless of the lattice arrangement. 
On the other hand, both F-F and Sc-Sc distances undergo a reduction in the structures that have not been experimentally resolved yet (see Figure~\ref{fgr:fingerprint}a). 
The smallest values are found in the monoclinic phase (4), where the Sc-Sc (F-F) separation becomes equal to 3.3~\AA{} (2.2~\AA{}).
These reductions are not unexpected given the relatively small unit-cell volume of this phase (see Figure~\ref{fgr:Dis} and Table~S1).
In the orthorhombic polymorph (3), a similar value of the F-F separation ($\sim$2.2~\AA{}) is accompanied by a slightly larger Sc-Sc distance (3.6~\AA{}).
In contrast, in the other orthorhombic phase (6), the interatomic separation Sc-Sc and F-F are the smallest, being 3.4~\AA{} and 2.8~\AA{}, respectively.
Finally, in the monoclinic structure (2), Sc-Sc and F-F distances are equal to 3.9~\AA{} and 2.6~\AA{}, respectively.

A more comprehensive interpretation of the interatomic distances in the considered \ce{ScF3} polymorphs can be obtained by adopting the F-fingerprint metric developed by Oganov and Valle\cite{ogan-vall09jcp}.
By discretizing the sum of the radial distribution function of each element pair, an $n$-dimensional vector is obtained, where $n$ is the number of discretized bins, and a quantitative similarity can be defined through the cosine-distance,
\begin{align}
	D_{cos} = \frac{1}{2} \left(1 - \frac{\vec{F_1}\cdot\vec{F_2}}{|\vec{F_1}| \,|\vec{F_2}|}\right),
	\label{eq:cos_dist}
\end{align}
where $F_1$ and $F_2$ are the radial distribution vectors of two different crystals.
Small values of $D_{cos}$ indicate close similarity, see Figure~\ref{fgr:fingerprint}b, where the results of Eq.~\eqref{eq:cos_dist} are plotted in a color matrix by contrasting each pair of structures among the six considered \ce{ScF3} polymorphs.
The similarity between the experimental low- and high-pressure phases (1 and 5) is evident from the plot.
Moreover, the resemblance between crystals (3) and (4) is highlighted, too.
The remaining two polymorphs, (2) and (6), are structurally closer to each other than to any other structure.
On the other hand, their similarity is less pronounced than for the other two pairs of crystals. 
Overall, it is noteworthy that the computationally predicted phases exhibit more remarkable structural differences among each other than the two experimentally resolved polymorphs.
 
%%%%%%%%%%%%%%%%%%%%%
\subsection{Energetic stability}
In the next step of our analysis, we assess the stability of the considered \ce{ScF3} crystal structures by calculating their formation energy per atom ($E_{form}$) according to the formula
\begin{equation}
  E_{form} = E(\text{ScF}_\text{3}) - \dfrac{1}{4} E(\text{Sc}) - \dfrac{3}{4} E(\text{F}), 
  \label{eq:form}
\end{equation}
where $E(x)$ is total energy per atom of compound $x$. 
For the elemental phases of F and Sc, the most stable experimentally crystal structure available in Materials Project\cite{Sc_phase,F_phase} has been taken.
We emphasize that the total energies entering Eq.~\eqref{eq:form} are computed from DFT assuming the system to be in a fixed geometry at 0~K.
As such, no thermodynamic effects are taken into account and the obtained values provide only a qualitative trend of the relative stability of the polymorphs.

The results shown in Table~\ref{tbl:formE} are consistent with the structural similarities discussed above.
The most stable structures are the low- and high-temperature polymorphs (1 and 5, respectively).
Interestingly, our calculations yield identical formation energies for these two materials.
Structures (3) and (4) exhibit very similar values of $E_{form}$, differing from each other only by 40~meV/atom, but being significantly less negative (by about 600~meV/atom) than those of polymorphs (1) and (5).
In contrast, structures (2) and (6) are characterized by formation energies equal to -4.19~eV/atom and -4.32~eV/atom, respectively.

We remark that this analysis holds for the single crystals modeled on the ideal structures sketched in Figure~\ref{fgr:Dis}. 
It is expected that the presence of defects, such as interstitial atoms and vacancies, will alter this picture.
A corresponding analysis is, however, beyond the scope of the present work. 

%%%%%%%%%%%%%%%%%%%%%%%%%%%%%%%%%
\subsection{Charge Distribution}

\begin{figure}
 \centering
 \includegraphics[width=0.5\textwidth]{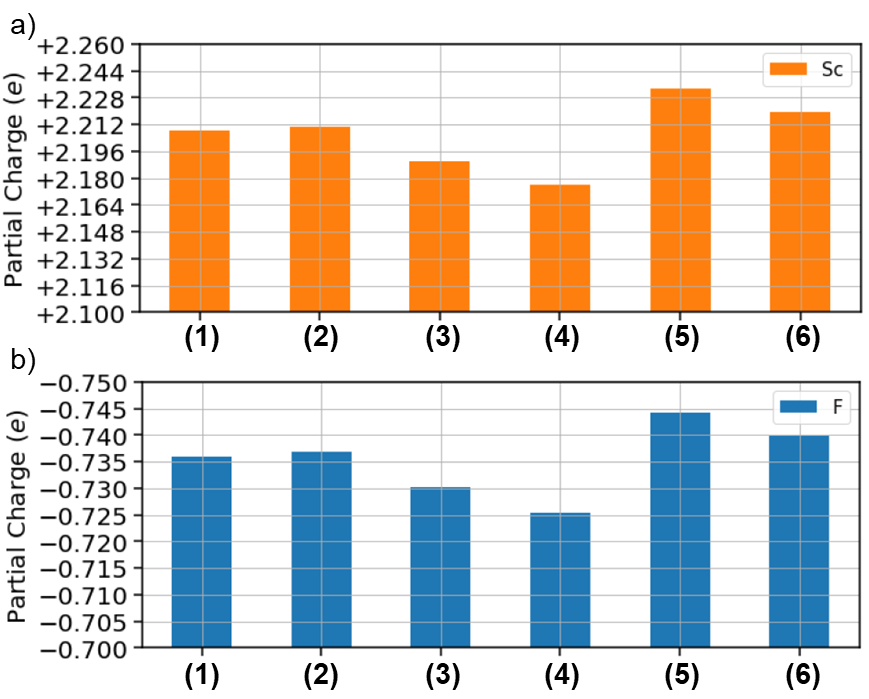}
 \caption{Mean partial charges calculated with the Bader scheme of a) Sc and b) F atoms in the six considered $\text{ScF}_\text{3}$ polymorphs.}
 \label{fgr:Pcharges}
\end{figure}

To analyze the chemical bonds in the different polymorphs, we perform partial charge analysis using the Bader scheme~\cite{henk+06cms}.
In agreement with the knowledge that \ce{ScF3} is an ionic crystal, in all considered phases, the Sc atom is positively charged and the F atoms are negatively charged (see Figure~\ref{fgr:Pcharges}).
Yet, the relative variations of the partial charges in the different polymorphs provide valuable information about the nature of the bonds: the larger the absolute values of the charges on each atomic species, the more ionic the bond.
In the experimental high-pressure \ce{ScF3} polymorph (5), the positive charge on Sc is about 2.33~$e$ and the negative charge on each F atom is almost as high as -0.75~$e$, thereby pointing to a high degree of ionicity in the bonds of this phase.
In contrast, the least pronounced ionicity among the considered polymorphs is found in the monoclinic crystal (4), where the charge on Sc (F) amounts to 2.17~$e$ (-0.72~$e$).
The other considered structures exhibit intermediate behaviors, with the orthorhombic crystal (6) having an Sc-F bond that is almost as ionic as the one in phase (5), and polymorph (3) displaying a more covalent bond similar to (4). 

From this analysis, it is evident that the similarities among the polymorphs based on interatomic distances and formation energies do not affect the partial charges. 
It should be noticed, though, that in absolute values the variations on the partial charges are small, ranging within an interval of less than 0.1~$e$.
This little spread is actually compatible with the negligible variations of the Sc-F distances among the polymorphs (see Figure~\ref{fgr:fingerprint}a).

%%%%%%%%%%%%%%%%%%%%%%%%%%%%%%%%%%%%%%%%%%
\subsection{Electronic Properties}

In the next step of this analysis, we inspect their electronic properties, focusing on the band gaps and on the density of states (DOS). 
With the exception of the high-pressure experimental polymorph (5), which exhibits a direct band gap at $\Gamma$ (see Table~\ref{tbl:formE}), all the other structures are characterized by an indirect band gap (see Figure~S2). 
The largest gaps are found for the known low- and high-pressure polymorphs (1 and 5) where $E_{gap}=6.12$~eV and $E_{gap}=6.14$~eV, respectively. 
The smallest direct gap obtained from DFT -- hereafter named \textit{optical} gap -- in the low-pressure polymorph (1) is equal to 6.64~eV.
Structures (2) and (6) are also wide-band-gap insulators with fundamental gaps as high as 5.28~eV and 5.77~eV, respectively, while the optical gaps amount to 5.52~eV and 5.81~eV, respectively (see Table~\ref{tbl:formE}).
Conversely, structures (3) and (4) are characterized by much smaller band gaps, around 2~eV.
Such a drastic reduction of the band-gap size in these polymorphs is a signature of their lower stability with respect to the experimental ones, as confirmed also by the values of formation energies reported in Table~\ref{tbl:formE}.
Overall, the values of the band gaps, both indirect and direct, follow the similarity trends found for the interatomic distances (Figure~\ref{fgr:fingerprint}b) and the formation energies (Table~\ref{tbl:formE}).

\begin{figure}
 \centering
 \includegraphics[width=0.465\textwidth]{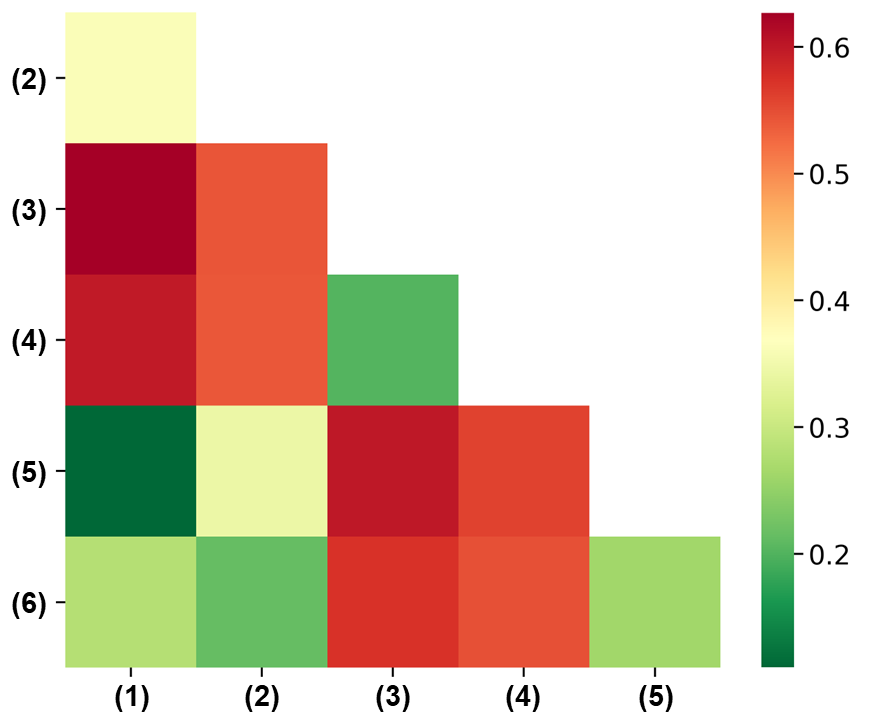}
 \caption{Similarity matrix based on the pair-wise cosine-distances evaluated from the F-fingerprint metric for the DOS computed for the six considered \ce{ScF3} polymorphs: Larger similarities correspond to low values (green) while smaller similarities to high values (red).}
 \label{fgr:FPrtDOS}
\end{figure}

The similarity among the considered polymorphs can be additionally evaluated for their DOS within an energy range between -15~eV and +15~eV around the Fermi energy.
The DOS for the six polymorphs is reported in the Supporting Information, Figure~S3. 
Here, we inspect the similarity matrix obtained by applying to these results the F-fingerprint metric~\cite{ogan-vall09jcp}.
In this case, some pre-processing of the data is necessary to evaluate Eq.~\eqref{eq:cos_dist}. 
First, all DOS are normalized with respect to the number of atoms in the unit cell and referenced to the energy value of the valence band maximum.
Then, the pairwise distance of the discretized functions is calculated and the cosine-distance is obtained.
The results plotted in Figure~\ref{fgr:FPrtDOS} confirm the similarities between pairs of structures seen in the structural analysis and in the formation energies: polymorphs (1) and (5) exhibit similar characteristics in the DOS; the same is true for structures (3) and (4) as well as for (2) and (6). 
Interestingly, the DOS of these two polymorphs, (2) and (6), is also quite similar to the one of structures (1) and (5).
The electronic structure of (2) and (3), instead, is quite different from that of all other polymorphs.
 
%%%%%%%%%%%%%%%%%%%%%%%%%%%%%
\subsection{X-ray spectroscopy}

In the second part of our study, we discuss the x-ray absorption spectra of \ce{ScF3} at the Sc and F K-edge. 
\bibnote{The Sc L$_{2,3}$-edge, although accessible in our experimental setup, cannot be simulated with sufficient accuracy in the adopted framework, given the available computational resources. 
As extensively discussed in Ref.~\citenum{vorw+17prb} with the example of CaO, a reliable convergence of the peak intensity of the two subedges in such weakly correlated systems requires a huge number of $|\mathbf{G}+\mathbf{q}|$ vectors for convergence, which we cannot afford at present.
For further details about the adopted formalism, see Refs.~\citenum{vorw+17prb,vorw+19es}.}
Results of first-principles calculations are reported on the six considered polymorphs as computed from the solution of the BSE (Eq.~\ref{eq:bse}) and as well as in the IPA.
By contrasting against each other the results obtained with these two methods, we can assess the role of excitonic effects, including quantifying exciton binding energies~\cite{cocc-drax15prb,cocc20pssrrl} and connecting the spectral fingerprints to the electronic structure of the materials.
Comparison with HERFD experimental data enables identifying the spectral fingerprints and assessing the composition of the sample in terms of the considered polymorphs.

%%%%%%%%%%%%%%%%%%%%%
\subsection{Sc K-edge Spectra}

\begin{figure*}
 \centering
 \includegraphics[width=\textwidth]{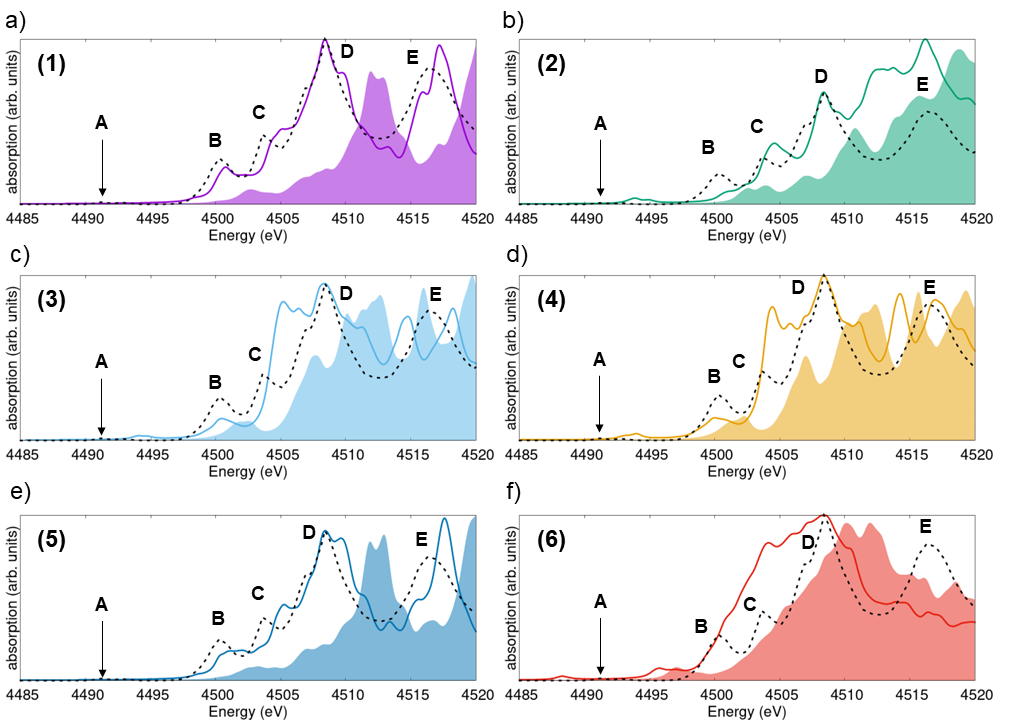}
 \caption{X-ray absorption spectra from the Sc K-edge calculated for a) polymorph (1), b) polymorph (2), c) polymorph (3), d) polymorph (4), e) polymorph (5), and f) polymorph (6). Solid lines (shaded areas) indicate results including (excluding) electron-hole interactions as computed from the solution of the BSE (in the IPA) with a Lorentzian broadening on 0.5~eV. The HERFD result (dashed line) is overlaid with all calculations and the main features therein are labeled in capital letters. The BSE spectra are aligned to the experimental reference with respect to the energy peak D.}
 \label{fgr:Sc-K}
\end{figure*}

We start from the x-ray absorption spectrum of \ce{ScF3} taken at the Sc K-edge, see Figure~\ref{fgr:Sc-K}.
The results calculated for each considered polymorph are reported on each panel and overlaid with the measurement (dashed line). 
First, let us consider the computational results obtained for the different structures (BSE: solid line; IPA: shaded area). 
All simulated spectra exhibit the typical shape from the K-edge of transition metals~\cite{degr+09jpcm}: a weak absorption onset is followed by stronger resonances at increasing energies.
Some phases are characterized by a pre-peak: it is found around 4493~eV in polymorphs (2), (3), and (4), while in structure (6) it appears at approximately 4487~eV.
Notably, such a feature is absent in the calculated spectra of polymorphs (1) and (5), representing the low- and high-pressure single-crystalline phases, respectively. 
A comparison among the computational results reveals analogous similarities to those highlighted in the structural properties, energetic stability, and charge distribution analysis reported above. 
Specifically, the spectra obtained for phases (3) and (4) resemble each other closely. 
Likewise, the spectra of polymorphs (1) and (5) are characterized by similar features.

It is worth reminding that the spectra discussed above are computed from the explicit evaluation of the matrix elements between core and conduction states (see Eq.~\ref{eq:epsilon}). Information about the chemical environment of the excited atoms is intrinsically encoded in the electronic structure obtained from DFT and further analysis on the local geometry of the targeted species is not requested by the adopted approach. For further details and comparisons among different methods for X-ray spectroscopy, we refer the readers to specialized reviews~\cite{besl20acr,degr+21jesrp}.

Excitonic effects are sizeable in all examined structures and manifest themselves mainly through redshifts of the peaks toward lower energies without a significant redistribution of the spectral weight among the absorption maxima.
The spectra of phases (3) and (4), however, do not exhibit this behavior, especially in the region between 4500 and 4515~eV.
Exciton binding energies, evaluated as the difference between excitation energies in the BSE solution and in the IPA~\cite{cocc-drax15prb,cocc+16prb,vorw+17prb,cocc20pssrrl}, are quite large in these spectra, on the order of a few eV and reaching values of approximately 5~eV for the most intense maxima in the spectra of polymorphs (1) and (5).
This result is not surprising, considering the ionic character of \ce{ScF3} and the consequent low screening therein. 
Similar values were obtained for core-level excitations in organic materials~\cite{cocc-drax15prb}.

\begin{figure}[h!]
 \centering
 \includegraphics[width=0.5\textwidth]{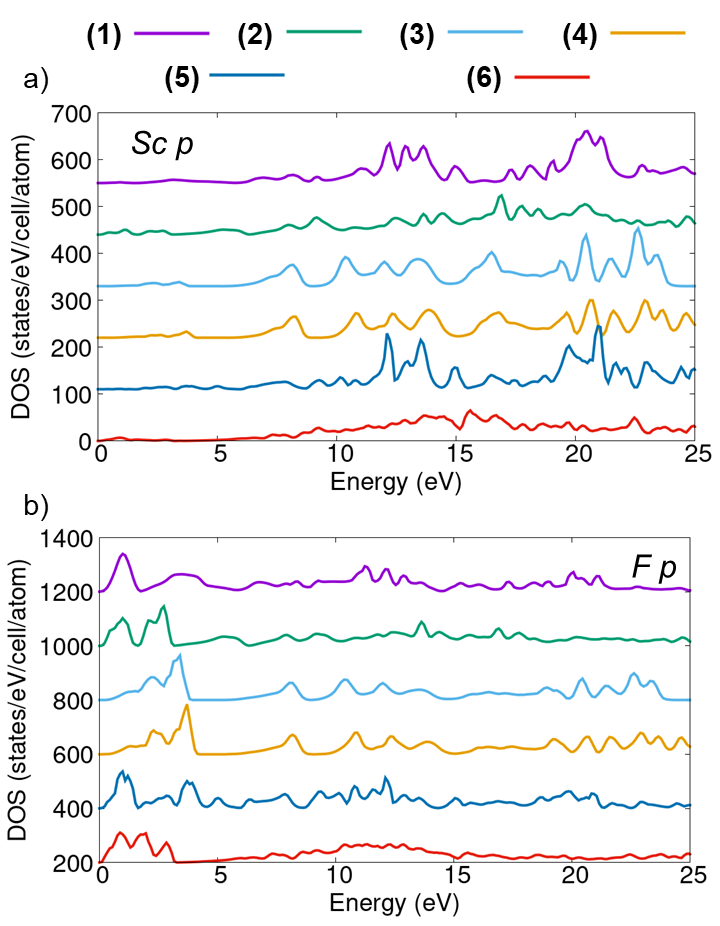}
 \caption{a) Sc and b) F contributions to the $p$-states in the projected density of states of the six \ce{ScF3} polymorphs considered in this work. The conduction band minimum is set at 0~eV.}
\label{fgr:PDOS-Sc+F}
\end{figure}

The spectra computed in the IPA (Figure~\ref{fgr:Sc-K}, shaded areas) can be closely connected with the Sc $p$-orbital contributions to the unoccupied PDOS (see Figure~\ref{fgr:PDOS-Sc+F}a), which represent the target states of these transitions.
The overall similarity between BSE and IPA spectra, except for the shifts of the resonances related to the excitonic effects discussed above, eases this analysis.
The strong resonances in the spectra around 10~eV above the (weak) onset are reflected in the PDOS where corresponding peaks are particularly pronounced in the energy region comprised between 10 and 15~eV (see Figure~\ref{fgr:PDOS-Sc+F}a). 
The similarities between the spectra of these low- and high-temperature polymorphs, (1) and (5), respectively, are reflected also in their PDOS, especially between 10 and 15~eV.
The maxima around 20~eV differ more substantially in these two structures.
Similarities are evident also between the Sc $p$-orbital contributions in the PDOS of phases (3) and (4), and, to a lesser extent due to the overall lower DOS magnitude, also for structures (2) and (6).
Careful inspection of Figure~\ref{fgr:PDOS-Sc+F}a reveals the presence of weak from the conduction band edge up to 5~eV in the PDOS of structures (2), (3), (4), and (6).
These states give rise to the pre-peak in the spectra of these polymorphs.
In contrast, no contributions from available Sc $p$-states are seen in the PDOS of the low- and high-temperature phases (1) and (5), in agreement with the absence of any corresponding pre-peaks in their spectra (see Figure~\ref{fgr:Sc-K}).

With the knowledge gained from this analysis, we can finally proceed with the comparison between the calculated spectra and the experimental dataset corresponding to the dashed line in 
all panels of Figure~\ref{fgr:Sc-K}. 
The experimental spectrum is corrected for overabsorption -- see details in the Supporting Information, Figure~S6.
Five main features are identified in the measurement: the weak pre-peak A, three maxima of increasing intensity building up the onset, B, C, and D, and the high-energy peak E. 
The pre-peak A can be associated with the low-energy feature in the spectra computed for polymorphs (2), (3), (4), and (6), although in all these cases the relative energy of this weak maximum is shifted with respect to the experimental reference.
Notice that the computational results are aligned with the energy of the most intense peak D.
The peaks at the onset, B, C, and D can be identified in all BSE spectra although the relative energies and oscillator strengths vary, thereby affecting the agreement with the experiment. 
The spectra calculated for polymorph (1) reproduces best all these characteristics (see Figure~\ref{fgr:Sc-K}a), including the relative oscillator strengths of the main peaks.
Good agreement with the experiment can be claimed also by the BSE result obtained for phase (5), where, in particular, the substructures of peak D are better captured than in the spectrum of phase (1). 
In the spectra computed for the other structures, we notice both an energy misalignment of the peaks and an incorrect description of their relative intensities.
All in all these findings suggest that the sample on which the HERFD measurement shown in Figure~\ref{fgr:Sc-K} was taken, contains predominantly phase (1) possibly mixed with polymorph (5). 
The missing pre-peak A in these two phases may point to the presence of additional, coexisting polymorphs where this feature is visible.
However, it can also be associated to defects or sample impurities that cannot be resolved by the calculations.

%%%%%%%%%%%%%%%%%%%%%%%%%%%%%%%%%%%%%
\subsection{F K-edge Spectra}

\begin{figure*}
 \centering
 \includegraphics[width=\textwidth]{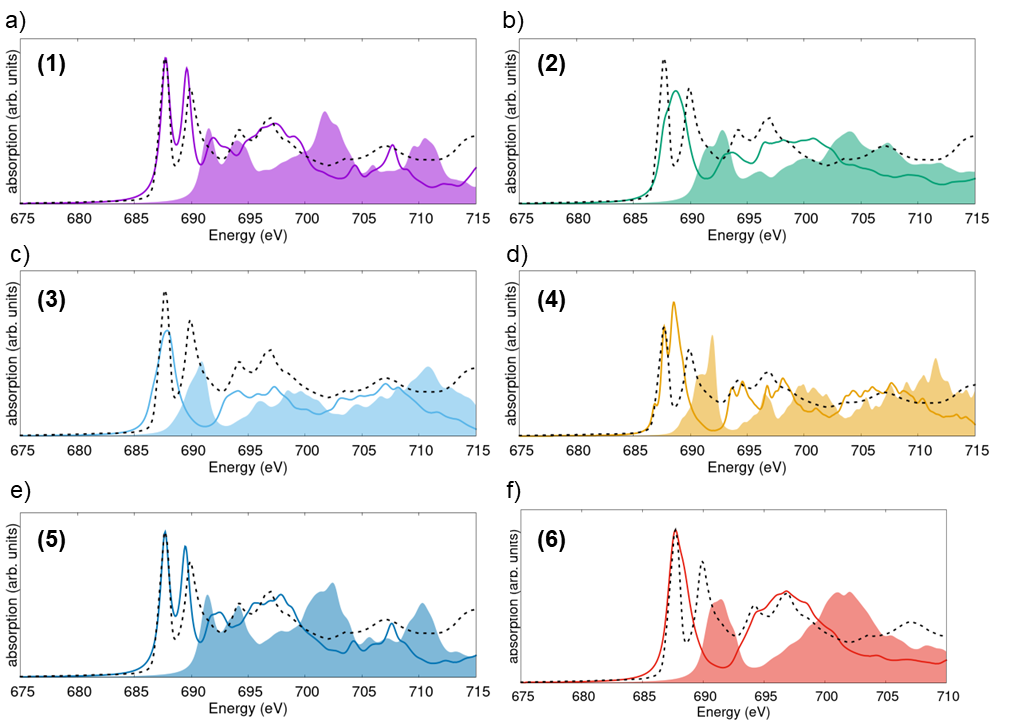}
 \caption{X-ray absorption spectra computed from the F K-edge of \ce{ScF3} in a) polymorph (1), b) polymorph (2), c) polymorph (3), d) polymorph (4), e) polymorph (5), and f) polymorph (6). Solid lines (shaded areas) indicate results including (excluding) electron-hole interactions as computed from the solution of the BSE (in the IPA). A Lorentzian broadening on 0.5~eV is applied to all spectra. The HERFD result (dashed line) is overlaid with all calculated spectra. The BSE spectra are aligned to the experimental reference at the energy of the peak at 687~eV.}
 \label{fgr:F-K}
\end{figure*}

Next, we examine the x-ray absorption spectra obtained from the F K-edge (see Figure~\ref{fgr:F-K}).
The results obtained for polymorphs (1) and (5), (3) and (4), as well as (2) and (6) exhibit mutually similar features.
In the spectra of structures (1) and (5), which correspond to the low- and high-temperature phases of \ce{ScF3}, two sharp maxima dominate the onset.
The presence of two bright excitations at the lowest energies is visible also in the spectra of phase (4), while in the spectra of polymorphs (2), (3), and (6), their energy separation is within the chosen broadening and, therefore, only one peak with a profile slightly deviating from an ideal Lorentzian is seen in Figures~\ref{fgr:F-K}b, \ref{fgr:F-K}c, and \ref{fgr:F-K}f.

Excitonic effects, evaluated again through the comparison between BSE and IPA results, are evident in all spectra.
In contrast to the Sc K-edge spectra where the explicit inclusion of electron-hole correlations in the calculations leads to an overall shift of the spectral weight toward lower energies and only to a slight redistribution of the oscillator strength (see Figure~\ref{fgr:Sc-K}), here, it enables the correct description of the sharp lowest-energy resonances (see Figure~\ref{fgr:F-K}).
In fact, although these peaks are present also in the IPA solution, their intensity is drastically underestimated due to the missing electron-hole interaction. 
This is a well-known shortcoming of this approximation~\cite{olov+09prb,olov+09jpcm,cocc+16prb,vorw+18jpcl,cocc20pssrrl}.
Quantitatively, exciton binding energies estimated for the F K-edge spectra of the considered \ce{ScF3} polymorphs are on the order of 3-4~eV, \textit{i.e.}, systematically larger than those obtained from the Sc K-edge.

The inspection of the PDOS shown in Figure~\ref{fgr:PDOS-Sc+F}b) helps explain the dominant spectral weight at low energies in these spectra. 
For all considered systems, a pronounced peak associated with the available F $p$-orbitals at the bottom of the conduction band is clearly visible for polymorphs (1) and (5).
In the PDOS computed for structures (2) and (6), available F $p$-states are spread over a broader energy window.
Finally, in the PDOS of polymorphs (3) and (4), the maxima for F $p$-orbitals are shifted at higher energies, between 2.5 and 4.5~eV.
The presence of F $p$-states at low energies in the unoccupied region follows chemical intuition: the empty orbitals of the 2$p$ shell of F lie low in energy due to the tendency of this element to accept electrons to complete the octet. 
The PDOS contributions displayed in Figure~\ref{fgr:PDOS-Sc+F}b support the similarity between the IPA spectra of polymorphs (1) and (5) as well as of phases (3) and (4).
Notably, for the latter two compounds, excitonic effects act differently, giving rise to a two-peak structure at the onset in (4) and to a single-peak structure in (3) in the BSE solutions.
In contrast, the similarity between the spectra of the experimentally stable structures (1) and (5) is reflected in both the IPA and the BSE solutions. 

The experimental data are very well reproduced by the results computed for polymorphs (1) and (5), see Figure~\ref{fgr:F-K}a and \ref{fgr:F-K}e. 
The two sharp resonances at the onset feature relative energies and intensities in close agreement with the measurements (notice that the computational results have been aligned in energy and intensity to the experimental ones with respect to the lowest-energy peak).
Higher-energy maxima are correctly reproduced, too.
The absence of the double-peak structure in the BSE spectra of the other polymorphs hinders the agreement with the experiment in this region.
However, especially for phase (3) and partly for phase (4), the higher-energy region (693-710~eV) is correctly reproduced by the calculations.
The analysis of the F K-edge spectra confirms the clear prevalence of polymorphs (1) and (5) in the measured sample, as anticipated from the study of the Sc K-edge spectra (see Figure~\ref{fgr:Sc-K}).
However, also from this analysis the presence of other polymorph in the probed sample cannot be totally excluded. 
In particular, the excellent agreement between the experimental spectral profile in the region 693-710~eV featured by the spectrum calculated for phase (3) -- even superior to the results obtained for polymorphs (1) and (5) in this energy range -- may suggest residues of this phase in the sample.
On the other hand, the presence of defects and impurities could also be responsible for these effects. 
Addressing this topic goes, however, beyond the scope of the present work.

%%%%%%%%%%%%%%%%%%%%%%%%%%%%%%%%%%%%%

\section{Summary and Conclusions}
In summary, we have performed a systematic first-principles analysis of six computationally predicted \ce{ScF3} polymorphs assessing their energetic stability, charge-density distribution, and electronic properties, aiming at identifying similarities between the various phases. 
Our results have revealed the analogy between the two experimentally resolved low- and high-temperature phases, indirectly confirming that the transition between them consists of a rigid lattice rotation.
The analysis of the x-ray absorption spectra of all structures sheds light on the electronic origin of the resonances and on the prominent excitonic effects characterizing them. 
The comparison against HERFD data reveals very good agreement with the results computed for the low- and high-temperature phases, confirming their predominance in the samples.
Some slight differences between the computed spectra of these systems and the experimental data raise some questions regarding the residual presence of other polymorphs, to date only computationally predicted, in the measured powder samples.
Yet, only a detailed analysis of the influence of defects and impurities on the spectra can support this hypothesis ruling out that such discrepancies are caused by these more common factors.

In conclusion, this work provides new insight into the structure-property relationships of \ce{ScF3} and its polymorphs, including both the experimentally resolved phases and the computationally predicted ones. Specifically, the information provided by the \textit{ab initio} characterization of the energetic stability, charge distribution, and electronic structure extends the existing knowledge of these systems. Further insight is given by the analysis of the X-ray absorption spectra by contrasting state-of-the-art measurements and calculations. We are confident that the results obtained for the computational phases will stimulate further experimental investigations on \ce{ScF3} single crystals.

\begin{acknowledgement}
The authors thank Ana Guilherme Buzanich for stimulating discussions in the initial stage of the project and the Paul Scherrer Institute for the provision of beamtime at the PHOENIX beamline of the Swiss Light Source (SLS).
This work was funded by the German Federal Ministry of Education and Research (Professorinnenprogramm III and grant nr. 03XP0328C), by the State of Lower Saxony (Professorinnen für Niedersachsen), and by the European Union through the Horizon 2020 research and innovation programs SCALE (H2020/ 2014–2020), grant agreement nr. 730105, and CALIPSOplus, grant agreement nr. 730872. 
The computational resources were provided by the North-German Supercomputing Alliance (HLRN), project nic00069, and by the high-performance computing cluster CARL at the University of Oldenburg, funded by the German Research Foundation (Project No. INST 184/157-1 FUGG) and by the Ministry of Science and Culture of the Lower Saxony State.

\end{acknowledgement}

\begin{suppinfo}
Crystallographic information about the simulated polymorphs, additional computational results (similarity matrix for the partial charges, band structure, and density of states of the considered polymorphs), and additional experimental data (X-ray diffraction patterns and resonant inelastic X-ray scattering map at the Sc K-edge) are reported.

\end{suppinfo}

%\bibliography{bib}

\begin{mcitethebibliography}{50}
\providecommand*\natexlab[1]{#1}
\providecommand*\mciteSetBstSublistMode[1]{}
\providecommand*\mciteSetBstMaxWidthForm[2]{}
\providecommand*\mciteBstWouldAddEndPuncttrue
  {\def\EndOfBibitem{\unskip.}}
\providecommand*\mciteBstWouldAddEndPunctfalse
  {\let\EndOfBibitem\relax}
\providecommand*\mciteSetBstMidEndSepPunct[3]{}
\providecommand*\mciteSetBstSublistLabelBeginEnd[3]{}
\providecommand*\EndOfBibitem{}
\mciteSetBstSublistMode{f}
\mciteSetBstMaxWidthForm{subitem}{(\alph{mcitesubitemcount})}
\mciteSetBstSublistLabelBeginEnd
  {\mcitemaxwidthsubitemform\space}
  {\relax}
  {\relax}

\bibitem[Ivers-Tiff{\'e}e \latin{et~al.}(2001)Ivers-Tiff{\'e}e, Weber, and
  Herbstritt]{iver+01jecs}
Ivers-Tiff{\'e}e,~E.; Weber,~A.; Herbstritt,~D. Materials and technologies for
  SOFC-components. \emph{J.~Eur.~Ceram.~Soc.} \textbf{2001}, \emph{21},
  1805--1811\relax
\mciteBstWouldAddEndPuncttrue
\mciteSetBstMidEndSepPunct{\mcitedefaultmidpunct}
{\mcitedefaultendpunct}{\mcitedefaultseppunct}\relax
\EndOfBibitem
\bibitem[Toropova \latin{et~al.}(2017)Toropova, Eskin, Kharakterova, and
  Dobatkina]{toro+17book}
Toropova,~L.; Eskin,~D.~G.; Kharakterova,~M.; Dobatkina,~T. \emph{Advanced
  aluminum alloys containing scandium: structure and properties}; Routledge,
  2017\relax
\mciteBstWouldAddEndPuncttrue
\mciteSetBstMidEndSepPunct{\mcitedefaultmidpunct}
{\mcitedefaultendpunct}{\mcitedefaultseppunct}\relax
\EndOfBibitem
\bibitem[Dorin \latin{et~al.}(2018)Dorin, Ramajayam, Vahid, and
  Langan]{dori+18book}
Dorin,~T.; Ramajayam,~M.; Vahid,~A.; Langan,~T. In \emph{Fundamentals of
  Aluminium Metallurgy}; Lumley,~R.~N., Ed.; Woodhead Publishing Series in
  Metals and Surface Engineering; Woodhead Publishing, 2018; pp 439--494\relax
\mciteBstWouldAddEndPuncttrue
\mciteSetBstMidEndSepPunct{\mcitedefaultmidpunct}
{\mcitedefaultendpunct}{\mcitedefaultseppunct}\relax
\EndOfBibitem
\bibitem[Grandfield(2021)]{grad21}
Grandfield,~J. 10-Year Outlook for the Global Scandium Market to 2031. 2021;
  \url{https://www.cmgroup.net/reports/the-10-year-outlook-for-the-global-scandium-market-to-2031/}\relax
\mciteBstWouldAddEndPuncttrue
\mciteSetBstMidEndSepPunct{\mcitedefaultmidpunct}
{\mcitedefaultendpunct}{\mcitedefaultseppunct}\relax
\EndOfBibitem
\bibitem[Petrakova \latin{et~al.}(2016)Petrakova, Panov, Gorbachev,
  Klimentenok, Perestoronin, Vishnyakov, and Anashkin]{petr+16book}
Petrakova,~O.~V.; Panov,~A.~V.; Gorbachev,~S.~N.; Klimentenok,~G.~N.;
  Perestoronin,~A.~V.; Vishnyakov,~S.~E.; Anashkin,~V.~S. In \emph{Light Metals
  2015}; Hyland,~M., Ed.; Springer International Publishing: Cham, 2016; pp
  93--96\relax
\mciteBstWouldAddEndPuncttrue
\mciteSetBstMidEndSepPunct{\mcitedefaultmidpunct}
{\mcitedefaultendpunct}{\mcitedefaultseppunct}\relax
\EndOfBibitem
\bibitem[Balomenos \latin{et~al.}(2021)Balomenos, Nazari, Davris, Abrenica,
  Pilihou, Mikeli, Panias, Patkar, and Xu]{balo+21book}
Balomenos,~E.; Nazari,~G.; Davris,~P.; Abrenica,~G.; Pilihou,~A.; Mikeli,~E.;
  Panias,~D.; Patkar,~S.; Xu,~W.-Q. \emph{Rare Metal Technology 2021};
  Springer, 2021; pp 217--228\relax
\mciteBstWouldAddEndPuncttrue
\mciteSetBstMidEndSepPunct{\mcitedefaultmidpunct}
{\mcitedefaultendpunct}{\mcitedefaultseppunct}\relax
\EndOfBibitem
\bibitem[Gentzmann \latin{et~al.}(2022)Gentzmann, Paul, Serrano, and
  Adam]{gent+22jge}
Gentzmann,~M.~C.; Paul,~A.; Serrano,~J.; Adam,~C. Understanding scandium
  leaching from bauxite residues of different geological backgrounds using
  statistical design of experiments. \emph{J.~Geochem.~Explor.} \textbf{2022},
  \emph{240}, 107041\relax
\mciteBstWouldAddEndPuncttrue
\mciteSetBstMidEndSepPunct{\mcitedefaultmidpunct}
{\mcitedefaultendpunct}{\mcitedefaultseppunct}\relax
\EndOfBibitem
\bibitem[Peters \latin{et~al.}(2019)Peters, Kaya, Dittrich, and
  Forsberg]{pete+19jsm}
Peters,~E.~M.; Kaya,~{\c{S}}.; Dittrich,~C.; Forsberg,~K. Recovery of scandium
  by crystallization techniques. \emph{J.~Sustain.~Metall.} \textbf{2019},
  \emph{5}, 48--56\relax
\mciteBstWouldAddEndPuncttrue
\mciteSetBstMidEndSepPunct{\mcitedefaultmidpunct}
{\mcitedefaultendpunct}{\mcitedefaultseppunct}\relax
\EndOfBibitem
\bibitem[Yagmurlu \latin{et~al.}(2021)Yagmurlu, Orberger, Dittrich, Croisé,
  Scharfenberg, Balomenos, Panias, Mikeli, Maier, Schneider, Friedrich,
  Dräger, Baumgärtner, Schmitz, Letmathe, Sakkas, Georgopoulos, and van~den
  Laan]{yagm+21mp}
Yagmurlu,~B.~; Orberger,~B.; Dittrich,~C.; Croise,~G.; Scharfenberg,~R,; Balomenos,~E.; Panias,~D.; Mikeli,~E.; Maier,~C.; Schneider,~R.; Friedrich,~B.; Draeger,~P.; Baumgaertner,~F.; Schmitz,~M.; Letmathe,~P.; Sakkas,~K.; Georgopoulos,~C.; van den Laan,~H. Sustainable Supply of Scandium for the EU
  Industries from Liquid Iron Chloride Based TiO2 Plants. \emph{Mater.~Proc.}
  \textbf{2021}, \emph{5}, 86, DOI: \doi{10.3390/materproc2021005086}\relax
\mciteBstWouldAddEndPuncttrue
\mciteSetBstMidEndSepPunct{\mcitedefaultmidpunct}
{\mcitedefaultendpunct}{\mcitedefaultseppunct}\relax
\EndOfBibitem
\bibitem[Greve \latin{et~al.}(2010)Greve, Martin, Lee, Chupas, Chapman, and
  Wilkinson]{grev+10jacs}
Greve,~B.~K.; Martin,~K.~L.; Lee,~P.~L.; Chupas,~P.~J.; Chapman,~K.~W.;
  Wilkinson,~A.~P. Pronounced negative thermal expansion from a simple
  structure: cubic ScF3. \emph{J.~Am.~Chem.~Soc.~} \textbf{2010}, \emph{132},
  15496--15498\relax
\mciteBstWouldAddEndPuncttrue
\mciteSetBstMidEndSepPunct{\mcitedefaultmidpunct}
{\mcitedefaultendpunct}{\mcitedefaultseppunct}\relax
\EndOfBibitem
\bibitem[Li \latin{et~al.}(2011)Li, Tang, Munoz, Keith, Tracy, Abernathy, and
  Fultz]{li+11prl}
Li,~C.~W.; Tang,~X.; Munoz,~J.~A.; Keith,~J.~B.; Tracy,~S.~J.;
  Abernathy,~D.~L.; Fultz,~B. Structural relationship between negative thermal
  expansion and quartic anharmonicity of cubic ScF 3. \emph{Phys.~Rev.~Lett.~}
  \textbf{2011}, \emph{107}, 195504\relax
\mciteBstWouldAddEndPuncttrue
\mciteSetBstMidEndSepPunct{\mcitedefaultmidpunct}
{\mcitedefaultendpunct}{\mcitedefaultseppunct}\relax
\EndOfBibitem
\bibitem[Lazar \latin{et~al.}(2015)Lazar, Bu{\v{c}}ko, and Hafner]{laza+15prb}
Lazar,~P.; Bu{\v{c}}ko,~T.; Hafner,~J. Negative thermal expansion of ScF 3:
  insights from density-functional molecular dynamics in the
  isothermal-isobaric ensemble. \emph{Phys.~Rev.~B} \textbf{2015}, \emph{92},
  224302\relax
\mciteBstWouldAddEndPuncttrue
\mciteSetBstMidEndSepPunct{\mcitedefaultmidpunct}
{\mcitedefaultendpunct}{\mcitedefaultseppunct}\relax
\EndOfBibitem
\bibitem[Piskunov \latin{et~al.}(2016)Piskunov, {\v{Z}}guns, Bocharov, Kuzmin,
  Purans, Kalinko, Evarestov, Ali, and Rocca]{pisk+16prb}
Piskunov,~S.; {\v{Z}}guns,~P.~A.; Bocharov,~D.; Kuzmin,~A.; Purans,~J.;
  Kalinko,~A.; Evarestov,~R.~A.; Ali,~S.~E.; Rocca,~F. Interpretation of
  unexpected behavior of infrared absorption spectra of ScF 3 beyond the
  quasiharmonic approximation. \emph{Phys.~Rev.~B} \textbf{2016}, \emph{93},
  214101\relax
\mciteBstWouldAddEndPuncttrue
\mciteSetBstMidEndSepPunct{\mcitedefaultmidpunct}
{\mcitedefaultendpunct}{\mcitedefaultseppunct}\relax
\EndOfBibitem
\bibitem[Attfield(2018)]{attf18fc}
Attfield,~J.~P. Mechanisms and Materials for NTE. \emph{Front.~Chem.}
  \textbf{2018}, 371\relax
\mciteBstWouldAddEndPuncttrue
\mciteSetBstMidEndSepPunct{\mcitedefaultmidpunct}
{\mcitedefaultendpunct}{\mcitedefaultseppunct}\relax
\EndOfBibitem
\bibitem[Oba \latin{et~al.}(2019)Oba, Tadano, Akashi, and Tsuneyuki]{oba+19prm}
Oba,~Y.; Tadano,~T.; Akashi,~R.; Tsuneyuki,~S. First-principles study of phonon
  anharmonicity and negative thermal expansion in ScF$_3$.
  \emph{Phys.~Rev.~Materials} \textbf{2019}, \emph{3}, 033601\relax
\mciteBstWouldAddEndPuncttrue
\mciteSetBstMidEndSepPunct{\mcitedefaultmidpunct}
{\mcitedefaultendpunct}{\mcitedefaultseppunct}\relax
\EndOfBibitem
\bibitem[Dove \latin{et~al.}(2020)Dove, Du, Wei, Keen, Tucker, and
  Phillips]{dove+20prb}
Dove,~M.~T.; Du,~J.; Wei,~Z.; Keen,~D.~A.; Tucker,~M.~G.; Phillips,~A.~E.
  Quantitative understanding of negative thermal expansion in scandium
  trifluoride from neutron total scattering measurements. \emph{Phys.~Rev.~B}
  \textbf{2020}, \emph{102}, 094105\relax
\mciteBstWouldAddEndPuncttrue
\mciteSetBstMidEndSepPunct{\mcitedefaultmidpunct}
{\mcitedefaultendpunct}{\mcitedefaultseppunct}\relax
\EndOfBibitem
\bibitem[Wei \latin{et~al.}(2020)Wei, Tan, Cai, Phillips, da~Silva, Kibble, and
  Dove]{wei+20prl}
Wei,~Z.; Tan,~L.; Cai,~G.; Phillips,~A.~E.; da~Silva,~I.; Kibble,~M.~G.;
  Dove,~M.~T. Colossal pressure-induced softening in scandium fluoride.
  \emph{Phys.~Rev.~Lett.~} \textbf{2020}, \emph{124}, 255502\relax
\mciteBstWouldAddEndPuncttrue
\mciteSetBstMidEndSepPunct{\mcitedefaultmidpunct}
{\mcitedefaultendpunct}{\mcitedefaultseppunct}\relax
\EndOfBibitem
\bibitem[Aleksandrov \latin{et~al.}(2002)Aleksandrov, Voronov, Vtyurin,
  Goryainov, Zamkova, Zinenko, and Krylov]{alek+02jetp}
Aleksandrov,~K.; Voronov,~V.; Vtyurin,~A.; Goryainov,~S.; Zamkova,~N.;
  Zinenko,~V.; Krylov,~A. Lattice dynamics and hydrostatic-pressure-induced
  phase transitions in ScF$_3$. \emph{J.~Exp.~Theo.~Phys.} \textbf{2002},
  \emph{94}, 977--984\relax
\mciteBstWouldAddEndPuncttrue
\mciteSetBstMidEndSepPunct{\mcitedefaultmidpunct}
{\mcitedefaultendpunct}{\mcitedefaultseppunct}\relax
\EndOfBibitem
\bibitem[Jain \latin{et~al.}(2013)Jain, Ong, Hautier, Chen, Richards, Dacek,
  Cholia, Gunter, Skinner, Ceder, and Persson]{jain+13aplm}
Jain,~A.; Ong,~S.~P.; Hautier,~G.; Chen,~W.; Richards,~W.~D.; Dacek,~S.;
  Cholia,~S.; Gunter,~D.; Skinner,~D.; Ceder,~G.; Persson,~K.~A. Commentary:
  The Materials Project: A materials genome approach to accelerating materials
  innovation. \emph{APL~Mater.} \textbf{2013}, \emph{1}, 011002\relax
\mciteBstWouldAddEndPuncttrue
\mciteSetBstMidEndSepPunct{\mcitedefaultmidpunct}
{\mcitedefaultendpunct}{\mcitedefaultseppunct}\relax
\EndOfBibitem
\bibitem[Giannozzi \latin{et~al.}(2017)Giannozzi, Andreussi, Brumme, Bunau,
  {Buongiorno Nardelli}, Calandra, Car, Cavazzoni, Ceresoli, Cococcioni,
  Colonna, Carnimeo, {Dal Corso}, de~Gironcoli, Delugas, DiStasio, Ferretti,
  Floris, Fratesi, Fugallo, Gebauer, Gerstmann, Giustino, Gorni, Jia, Kawamura,
  Ko, Kokalj, K{\"u}{\c{c}}{\"u}kbenli, Lazzeri, Marsili, Marzari, Mauri,
  Nguyen, Nguyen, Otero-de-la Roza, Paulatto, Ponc{\'e}, Rocca, Sabatini,
  Santra, Schlipf, Seitsonen, Smogunov, Timrov, Thonhauser, Umari, Vast, Wu,
  and Baroni]{gian+17jpcm}
Giannozzi, P.; Andreussi, O.; Brumme, T.; Bunau, O.; {Buongiorno Nardelli}, M.; Calandra, M.; Car, R.; Cavazzoni, C.; Ceresoli, D.; Cococcioni, M.; Colonna, N.; Carnimeo, I.; {Dal Corso}, A.; de Gironcoli, S.; Delugas, P.; DiStasio, R. A.; Ferretti, A.; Floris, A.; Fratesi, G.; Fugallo, G.; Gebauer, R.; Gerstmann, U.; Giustino, F.; Gorni, T.; Jia, J.; Kawamura, M.; Ko, H-Y; Kokalj, A.; K{\"u}{\c{c}}{\"u}kbenli, E.; Lazzeri, M.; Marsili, M.; Marzari, N.; Mauri, F.; Nguyen, N. L.; Nguyen, H-V; Otero-de-la-Roza, A.; Paulatto, L.; Ponc{\'e}, S.; Rocca, D.; Sabatini, R.; Santra, B.; Schlipf, M.; Seitsonen, A. P.; Smogunov, A.; Timrov, I.; Thonhauser, T.; Umari, P.; Vast, N.; Wu, X.; Baroni, S. Advanced capabilities for materials modelling
  with Quantum ESPRESSO. \emph{J.~Phys.:~Condens.~Matter.~} \textbf{2017},
  \emph{29}, 465901\relax
\mciteBstWouldAddEndPuncttrue
\mciteSetBstMidEndSepPunct{\mcitedefaultmidpunct}
{\mcitedefaultendpunct}{\mcitedefaultseppunct}\relax
\EndOfBibitem
\bibitem[{Dal Corso}(2014)]{dalc14cms}
{Dal Corso},~A. Pseudopotentials periodic table: From H to Pu.
  \emph{Comp.~Mater.~Sci.~} \textbf{2014}, \emph{95}, 337--350\relax
\mciteBstWouldAddEndPuncttrue
\mciteSetBstMidEndSepPunct{\mcitedefaultmidpunct}
{\mcitedefaultendpunct}{\mcitedefaultseppunct}\relax
\EndOfBibitem
\bibitem[Perdew \latin{et~al.}(1996)Perdew, Burke, and Ernzerhof]{perd+96prl}
Perdew,~J.~P.; Burke,~K.; Ernzerhof,~M. Generalized Gradient Approximation Made
  Simple. \emph{Phys.~Rev.~Lett.~} \textbf{1996}, \emph{77}, 3865--3868\relax
\mciteBstWouldAddEndPuncttrue
\mciteSetBstMidEndSepPunct{\mcitedefaultmidpunct}
{\mcitedefaultendpunct}{\mcitedefaultseppunct}\relax
\EndOfBibitem
\bibitem[Henkelman \latin{et~al.}(2006)Henkelman, Arnaldsson, and
  Jónsson]{henk+06cms}
Henkelman,~G.; Arnaldsson,~A.; Jónsson,~H. A fast and robust algorithm for
  Bader decomposition of charge density. \emph{Comp.~Mater.~Sci.~}
  \textbf{2006}, \emph{36}, 354--360\relax
\mciteBstWouldAddEndPuncttrue
\mciteSetBstMidEndSepPunct{\mcitedefaultmidpunct}
{\mcitedefaultendpunct}{\mcitedefaultseppunct}\relax
\EndOfBibitem
\bibitem[Sanville \latin{et~al.}(2007)Sanville, Kenny, Smith, and
  Henkelman]{sanv+07jcc}
Sanville,~E.; Kenny,~S.~D.; Smith,~R.; Henkelman,~G. Improved grid-based
  algorithm for Bader charge allocation. \emph{J.~Comput.~Chem.~}
  \textbf{2007}, \emph{28}, 899--908\relax
\mciteBstWouldAddEndPuncttrue
\mciteSetBstMidEndSepPunct{\mcitedefaultmidpunct}
{\mcitedefaultendpunct}{\mcitedefaultseppunct}\relax
\EndOfBibitem
\bibitem[Tang \latin{et~al.}(2009)Tang, Sanville, and Henkelman]{tang+09jpcm}
Tang,~W.; Sanville,~E.; Henkelman,~G. A grid-based Bader analysis algorithm
  without lattice bias. \emph{J.~Phys.:~Condens.~Matter.~} \textbf{2009},
  \emph{21}, 084204\relax
\mciteBstWouldAddEndPuncttrue
\mciteSetBstMidEndSepPunct{\mcitedefaultmidpunct}
{\mcitedefaultendpunct}{\mcitedefaultseppunct}\relax
\EndOfBibitem
\bibitem[Gulans \latin{et~al.}(2014)Gulans, Kontur, Meisenbichler, Nabok,
  Pavone, Rigamonti, Sagmeister, Werner, and Draxl]{gula+14jpcm}
Gulans,~A.; Kontur,~S.; Meisenbichler,~C.; Nabok,~D.; Pavone,~P.;
  Rigamonti,~S.; Sagmeister,~S.; Werner,~U.; Draxl,~C. exciting: a
  full-potential all-electron package implementing density-functional theory
  and many-body perturbation theory. \emph{J.~Phys.:~Condens.~Matter.~}
  \textbf{2014}, \emph{26}, 363202\relax
\mciteBstWouldAddEndPuncttrue
\mciteSetBstMidEndSepPunct{\mcitedefaultmidpunct}
{\mcitedefaultendpunct}{\mcitedefaultseppunct}\relax
\EndOfBibitem
\bibitem[Vorwerk \latin{et~al.}(2017)Vorwerk, Cocchi, and Draxl]{vorw+17prb}
Vorwerk,~C.; Cocchi,~C.; Draxl,~C. Addressing electron-hole correlation in core
  excitations of solids: An all-electron many-body approach from first
  principles. \emph{Phys.~Rev.~B} \textbf{2017}, \emph{95}, 155121\relax
\mciteBstWouldAddEndPuncttrue
\mciteSetBstMidEndSepPunct{\mcitedefaultmidpunct}
{\mcitedefaultendpunct}{\mcitedefaultseppunct}\relax
\EndOfBibitem
\bibitem[Vorwerk \latin{et~al.}(2019)Vorwerk, Aurich, Cocchi, and
  Draxl]{vorw+19es}
Vorwerk,~C.; Aurich,~B.; Cocchi,~C.; Draxl,~C. Bethe--Salpeter equation for
  absorption and scattering spectroscopy: implementation in the exciting code.
  \emph{Electr.~Struct.} \textbf{2019}, \emph{1}, 037001\relax
\mciteBstWouldAddEndPuncttrue
\mciteSetBstMidEndSepPunct{\mcitedefaultmidpunct}
{\mcitedefaultendpunct}{\mcitedefaultseppunct}\relax
\EndOfBibitem
\bibitem[Salpeter and Bethe(1951)Salpeter, and Bethe]{salp-beth51pr}
Salpeter,~E.~E.; Bethe,~H.~A. A relativistic equation for bound-state problems.
  \emph{Phys.~Rev.~} \textbf{1951}, \emph{84}, 1232\relax
\mciteBstWouldAddEndPuncttrue
\mciteSetBstMidEndSepPunct{\mcitedefaultmidpunct}
{\mcitedefaultendpunct}{\mcitedefaultseppunct}\relax
\EndOfBibitem
\bibitem[Rohlfing and Louie(2000)Rohlfing, and Louie]{rohl-loui00prb}
Rohlfing,~M.; Louie,~S.~G. Electron-hole excitations and optical spectra from
  first principles. \emph{Phys.~Rev.~B} \textbf{2000}, \emph{62},
  4927--4944\relax
\mciteBstWouldAddEndPuncttrue
\mciteSetBstMidEndSepPunct{\mcitedefaultmidpunct}
{\mcitedefaultendpunct}{\mcitedefaultseppunct}\relax
\EndOfBibitem
\bibitem[Puschnig and Ambrosch-Draxl(2002)Puschnig, and
  Ambrosch-Draxl]{pusc-ambr02prb}
Puschnig,~P.; Ambrosch-Draxl,~C. Optical absorption spectra of semiconductors
  and insulators including electron-hole correlations: An ab initio study
  within the LAPW method. \emph{Phys.~Rev.~B} \textbf{2002}, \emph{66},
  165105\relax
\mciteBstWouldAddEndPuncttrue
\mciteSetBstMidEndSepPunct{\mcitedefaultmidpunct}
{\mcitedefaultendpunct}{\mcitedefaultseppunct}\relax
\EndOfBibitem
\bibitem[Onida \latin{et~al.}(2002)Onida, Reining, and Rubio]{onid+02rmp}
Onida,~G.; Reining,~L.; Rubio,~A. Electronic excitations: density-functional
  versus many-body Green’s-function approaches. \emph{Rev.~Mod.~Phys.~}
  \textbf{2002}, \emph{74}, 601\relax
\mciteBstWouldAddEndPuncttrue
\mciteSetBstMidEndSepPunct{\mcitedefaultmidpunct}
{\mcitedefaultendpunct}{\mcitedefaultseppunct}\relax
\EndOfBibitem
\bibitem[Gentzmann \latin{et~al.}(2021)Gentzmann, Schraut, Vogel, G{\"a}bler,
  Huthwelker, and Adam]{gent+21ag}
Gentzmann,~M.~C.; Schraut,~K.; Vogel,~C.; G{\"a}bler,~H.-E.; Huthwelker,~T.;
  Adam,~C. Investigation of scandium in bauxite residues of different origin.
  \emph{Appl.~Geochem.} \textbf{2021}, \emph{126}, 104898\relax
\mciteBstWouldAddEndPuncttrue
\mciteSetBstMidEndSepPunct{\mcitedefaultmidpunct}
{\mcitedefaultendpunct}{\mcitedefaultseppunct}\relax
\EndOfBibitem
\bibitem[Ramilli \latin{et~al.}(2017)Ramilli, Bergamaschi, Andrae,
  Br{\"u}ckner, Cartier, Dinapoli, Fr{\"o}jdh, Greiffenberg, Hutwelker,
  Lopez-Cuenca, Mezza, Mozzanica, Ruat, Redford, Schmitt, Shi, Tinti, and
  Zhang]{rami+17jinst}
Ramilli, M.; Bergamaschi, Andrae, M.; Br{\"u}ckner, M.; Cartier, S.; Dinapoli, R.; Fr{\"o}jdh, E.; Greiffenberg, D.; Hutwelker, T.; Lopez-Cuenca, C.; Mezza, D.; Mozzanica, A.; Ruat, M.; Redford, S.; Schmitt, B.; Shi, X.; Tinti, G.; Zhang, J.  Measurements with M{\"O}NCH, a 25 $\mu$m pixel
  pitch hybrid pixel detector. \emph{J.~Instrum.} \textbf{2017}, \emph{12},
  C01071\relax
\mciteBstWouldAddEndPuncttrue
\mciteSetBstMidEndSepPunct{\mcitedefaultmidpunct}
{\mcitedefaultendpunct}{\mcitedefaultseppunct}\relax
\EndOfBibitem
\bibitem[Piamonteze \latin{et~al.}(2012)Piamonteze, Flechsig, Rusponi, Dreiser,
  Heidler, Schmidt, Wetter, Calvi, Schmidt, Pruchova, Krempasky, Quitmann,
  Brune, and Nolting]{piam+12jsr}
Piamonteze,~C.; Flechsig,~U.; Rusponi,~S.; Dreiser,~J.; Heidler,~J.;
  Schmidt,~M.; Wetter,~R.; Calvi,~M.; Schmidt,~T.; Pruchova,~H.; Krempasky,~J.;
  Quitmann,~C.; Brune,~H.; Nolting,~F. X-Treme beamline at SLS: X-ray magnetic
  circular and linear dichroism at high field and low temperature.
  \emph{J.~Synchrotron~Radiat.} \textbf{2012}, \emph{19}, 661--674\relax
\mciteBstWouldAddEndPuncttrue
\mciteSetBstMidEndSepPunct{\mcitedefaultmidpunct}
{\mcitedefaultendpunct}{\mcitedefaultseppunct}\relax
\EndOfBibitem
\bibitem[Momma and Izumi(2011)Momma, and Izumi]{momm-izum11jacr}
Momma,~K.; Izumi,~F. VESTA 3 for three-dimensional visualization of crystal,
  volumetric and morphology data. \emph{J.~Appl.~Cryst.~} \textbf{2011},
  \emph{44}, 1272--1276\relax
\mciteBstWouldAddEndPuncttrue
\mciteSetBstMidEndSepPunct{\mcitedefaultmidpunct}
{\mcitedefaultendpunct}{\mcitedefaultseppunct}\relax
\EndOfBibitem
\bibitem[Oganov and Valle(2009)Oganov, and Valle]{ogan-vall09jcp}
Oganov,~A.~R.; Valle,~M. How to quantify energy landscapes of solids.
  \emph{J.~Chem.~Phys.~} \textbf{2009}, \emph{130}, 104504\relax
\mciteBstWouldAddEndPuncttrue
\mciteSetBstMidEndSepPunct{\mcitedefaultmidpunct}
{\mcitedefaultendpunct}{\mcitedefaultseppunct}\relax
\EndOfBibitem
\bibitem[Persson(2016)]{Sc_phase}
Persson,~K. Materials Data on Sc (SG:194) by Materials Project. 2016\relax
\mciteBstWouldAddEndPuncttrue
\mciteSetBstMidEndSepPunct{\mcitedefaultmidpunct}
{\mcitedefaultendpunct}{\mcitedefaultseppunct}\relax
\EndOfBibitem
\bibitem[Persson(2016)]{F_phase}
Persson,~K. Materials Data on F2 (SG:15) by Materials Project. 2016\relax
\mciteBstWouldAddEndPuncttrue
\mciteSetBstMidEndSepPunct{\mcitedefaultmidpunct}
{\mcitedefaultendpunct}{\mcitedefaultseppunct}\relax
\EndOfBibitem
\bibitem[Not()]{Note-1}
The Sc L$_{2,3}$-edge, although accessible in our experimental setup, cannot be
  simulated with sufficient accuracy in the adopted framework, given the
  available computational resources. As extensively discussed in
  Ref.~\citenum{vorw+17prb} with the example of CaO, a reliable convergence of
  the peak intensity of the two subedges in such weakly correlated systems
  requires a huge number of $|\mathbf{G}+\mathbf{q}|$ vectors for convergence,
  which we cannot afford at present. For further details about the adopted
  formalism, see Refs.~\citenum{vorw+17prb,vorw+19es}.\relax
\mciteBstWouldAddEndPunctfalse
\mciteSetBstMidEndSepPunct{\mcitedefaultmidpunct}
{}{\mcitedefaultseppunct}\relax
\EndOfBibitem
\bibitem[Cocchi and Draxl(2015)Cocchi, and Draxl]{cocc-drax15prb}
Cocchi,~C.; Draxl,~C. Bound excitons and many-body effects in x-ray absorption
  spectra of azobenzene-functionalized self-assembled monolayers.
  \emph{Phys.~Rev.~B} \textbf{2015}, \emph{92}, 205105\relax
\mciteBstWouldAddEndPuncttrue
\mciteSetBstMidEndSepPunct{\mcitedefaultmidpunct}
{\mcitedefaultendpunct}{\mcitedefaultseppunct}\relax
\EndOfBibitem
\bibitem[Cocchi(2020)]{cocc20pssrrl}
Cocchi,~C. X-Ray Absorption Fingerprints from Cs Atoms in Cs$_3$Sb.
  \emph{Phys.~Status~Solidi~(RRL)} \textbf{2020}, \emph{14}, 2000194\relax
\mciteBstWouldAddEndPuncttrue
\mciteSetBstMidEndSepPunct{\mcitedefaultmidpunct}
{\mcitedefaultendpunct}{\mcitedefaultseppunct}\relax
\EndOfBibitem
\bibitem[De~Groot \latin{et~al.}(2009)De~Groot, Vank{\'o}, and
  Glatzel]{degr+09jpcm}
De~Groot,~F.; Vank{\'o},~G.; Glatzel,~P. The 1s x-ray absorption pre-edge
  structures in transition metal oxides. \emph{J.~Phys.:~Condens.~Matter.~}
  \textbf{2009}, \emph{21}, 104207\relax
\mciteBstWouldAddEndPuncttrue
\mciteSetBstMidEndSepPunct{\mcitedefaultmidpunct}
{\mcitedefaultendpunct}{\mcitedefaultseppunct}\relax
\EndOfBibitem
\bibitem[Besley(2020)]{besl20acr}
Besley,~N.~A. Density functional theory based methods for the calculation of
  X-ray spectroscopy. \emph{Acc.~Chem.~Res.~} \textbf{2020}, \emph{53},
  1306--1315\relax
\mciteBstWouldAddEndPuncttrue
\mciteSetBstMidEndSepPunct{\mcitedefaultmidpunct}
{\mcitedefaultendpunct}{\mcitedefaultseppunct}\relax
\EndOfBibitem
\bibitem[{de Groot} \latin{et~al.}(2021){de Groot}, Elnaggar, Frati, pan Wang,
  Delgado-Jaime, {van Veenendaal}, Fernandez-Rodriguez, Haverkort, Green, {van
  der Laan}, Kvashnin, Hariki, Ikeno, Ramanantoanina, Daul, Delley, Odelius,
  Lundberg, Kuhn, Bokarev, Shirley, Vinson, Gilmore, Stener, Fronzoni, Decleva,
  Kruger, Retegan, Joly, Vorwerk, Draxl, Rehr, and Tanaka]{degr+21jesrp}
%{de Groot},~F.~M. \latin{et~al.}  
De~Groot,~F.; Elnaggar,~H.; Frati,~F.; Wang,~R.; Delgado-Jaime,~M.~U.; Van~Veenendaal, M.; Fernandez-Rodriguez,~J.; Haverkort,~M.~W.; Green,~R.J.; van~der~Laan,~G.; Kvashnin,~Y.; Hariki,~A.; Ikeno,~H.; Ramanantoanina,~H.; Daul,~C.; Delley,~B; Odelius,~M.; Lundberg,~M.; Kuhn,~O.; Bokarev,~S.~I.; Shirley,~E.; Vinson,~J.; Gilmore,~K.; Stener,~M.; Fronzoni,~G.; Decleva,~P.; Kruger,~P.; Retegan,~M.; Joly,~Y.; Vorwerk,~C.; Draxl,~C.; Rehr,~J.; Tanaka~A. 2p x-ray absorption spectroscopy of 3d
  transition metal systems. \emph{J.~Electron~Spectrosc.~Relat.~Phenom.~}
  \textbf{2021}, \emph{249}, 147061\relax
\mciteBstWouldAddEndPuncttrue
\mciteSetBstMidEndSepPunct{\mcitedefaultmidpunct}
{\mcitedefaultendpunct}{\mcitedefaultseppunct}\relax
\EndOfBibitem
\bibitem[Cocchi \latin{et~al.}(2016)Cocchi, Zschiesche, Nabok, Mogilatenko,
  Albrecht, Galazka, Kirmse, Draxl, and Koch]{cocc+16prb}
Cocchi,~C.; Zschiesche,~H.; Nabok,~D.; Mogilatenko,~A.; Albrecht,~M.;
  Galazka,~Z.; Kirmse,~H.; Draxl,~C.; Koch,~C.~T. Atomic signatures of local
  environment from core-level spectroscopy in
  $\ensuremath{\beta}\text{\ensuremath{-}}{\mathrm{Ga}}_{2}{\mathrm{O}}_{3}$.
  \emph{Phys.~Rev.~B} \textbf{2016}, \emph{94}, 075147\relax
\mciteBstWouldAddEndPuncttrue
\mciteSetBstMidEndSepPunct{\mcitedefaultmidpunct}
{\mcitedefaultendpunct}{\mcitedefaultseppunct}\relax
\EndOfBibitem
\bibitem[Olovsson \latin{et~al.}(2009)Olovsson, Tanaka, Mizoguchi, Puschnig,
  and Ambrosch-Draxl]{olov+09prb}
Olovsson,~W.; Tanaka,~I.; Mizoguchi,~T.; Puschnig,~P.; Ambrosch-Draxl,~C.
  All-electron Bethe-Salpeter calculations for shallow-core x-ray absorption
  near-edge structures. \emph{Phys.~Rev.~B} \textbf{2009}, \emph{79},
  041102\relax
\mciteBstWouldAddEndPuncttrue
\mciteSetBstMidEndSepPunct{\mcitedefaultmidpunct}
{\mcitedefaultendpunct}{\mcitedefaultseppunct}\relax
\EndOfBibitem
\bibitem[Olovsson \latin{et~al.}(2009)Olovsson, Tanaka, Puschnig, and
  Ambrosch-Draxl]{olov+09jpcm}
Olovsson,~W.; Tanaka,~I.; Puschnig,~P.; Ambrosch-Draxl,~C. Near-edge structures
  from first principles all-electron Bethe--Salpeter equation calculations.
  \emph{J.~Phys.:~Condens.~Matter.~} \textbf{2009}, \emph{21}, 104205\relax
\mciteBstWouldAddEndPuncttrue
\mciteSetBstMidEndSepPunct{\mcitedefaultmidpunct}
{\mcitedefaultendpunct}{\mcitedefaultseppunct}\relax
\EndOfBibitem
\bibitem[Vorwerk \latin{et~al.}(2018)Vorwerk, Hartmann, Cocchi, Sadoughi,
  Habisreutinger, F{\'e}lix, Wilks, Snaith, Bär, and Draxl]{vorw+18jpcl}
Vorwerk,~C.; Hartmann,~C.; Cocchi,~C.; Sadoughi,~G.; Habisreutinger,~S.~N.;
  F{\'e}lix,~R.; Wilks,~R.~G.; Snaith,~H.~J.; Baer,~M.; Draxl,~C.
  Exciton-dominated core-level absorption spectra of hybrid organic--inorganic
  lead halide perovskites. \emph{J.~Phys.~Chem.~Lett.} \textbf{2018}, \emph{9},
  1852--1858\relax
\mciteBstWouldAddEndPuncttrue
\mciteSetBstMidEndSepPunct{\mcitedefaultmidpunct}
{\mcitedefaultendpunct}{\mcitedefaultseppunct}\relax
\EndOfBibitem
\end{mcitethebibliography}

\providecommand{\latin}[1]{#1}
\makeatletter
\providecommand{\doi}
  {\begingroup\let\do\@makeother\dospecials
  \catcode`\{=1 \catcode`\}=2 \doi@aux}
\providecommand{\doi@aux}[1]{\endgroup\texttt{#1}}
\makeatother
\providecommand*\mcitethebibliography{\thebibliography}
\csname @ifundefined\endcsname{endmcitethebibliography}
  {\let\endmcitethebibliography\endthebibliography}{}

%%%%%%%%%%%%%%%%%%%%%%%%
\newpage

\centering
\section*{For Table of Contents Only}

\begin{figure*}
 \centering
\includegraphics[height=4.45 cm]{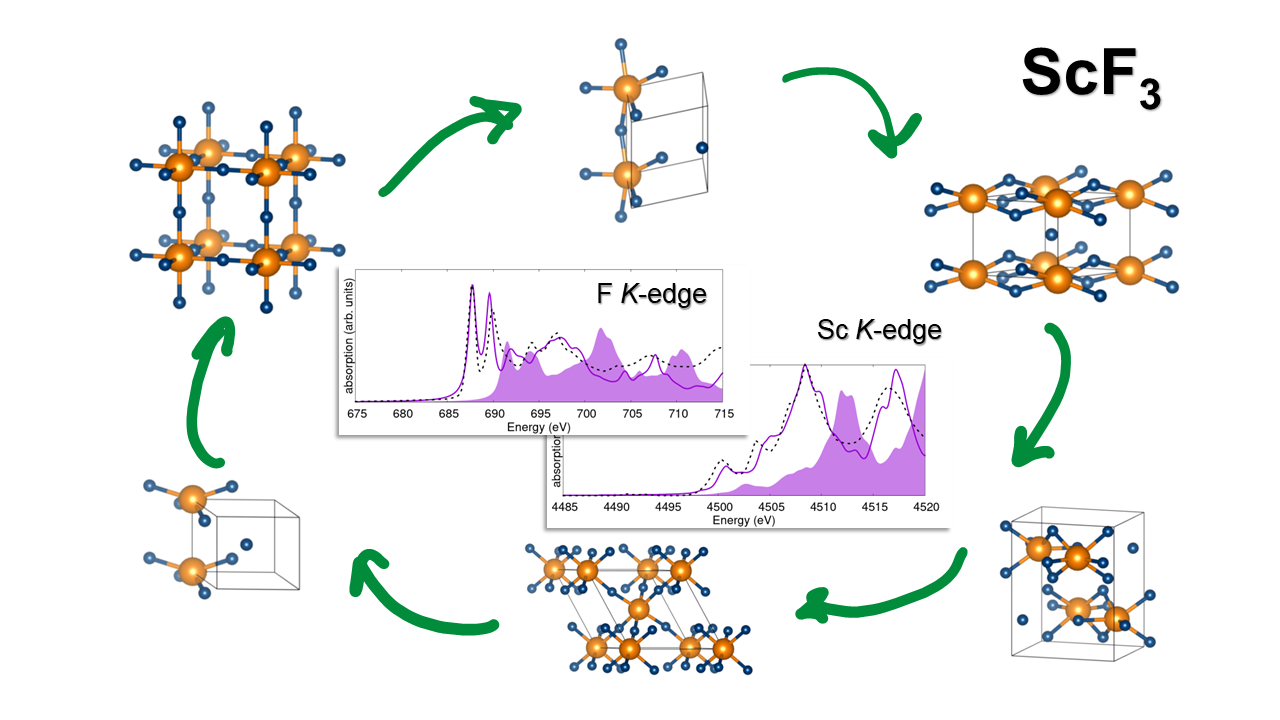}
\end{figure*}

The stability and electronic structure of six \ce{ScF3} crystalline polymorphs are calculated from first principles using density-functional theory. With the aid of X-ray spectroscopy measurements on a powder sample, the fingerprints of the material from the Sc and F K-edges are identified and contrasted against the solution of the Bethe-Salpeter equation on the computationally predicted structures, which enables the assessment of excitonic effects in the spectra.
%%%%%%%%%%%%%%%%%%%%%%%%%%

\end{document}